\documentclass[pra,twocolumn,preprintnumbers,amsmath,amssymb,superscriptaddress,showpacs,longbibliography]{revtex4-2}
\usepackage{tabularx,amsmath,amsfonts,amssymb,boxedminipage,graphicx,mathtools,appendix,physics}
\usepackage{epstopdf, float,siunitx,makecell}
\usepackage{algorithm2e}
\usepackage{subfigure}
\usepackage[final]{hyperref}
\usepackage{csquotes,booktabs}
\MakeOuterQuote{"}
\RestyleAlgo{boxruled}
\hypersetup{
	colorlinks=true,       
	linkcolor=blue,          
	citecolor=blue,        
	filecolor=magenta,      
	urlcolor=blue         
}

\allowdisplaybreaks  
\makeatletter
\@ifundefined{widetext}{ }{}
\makeatother

\newcommand{\Kcal}{\mathcal{K}}

\newenvironment{customthm}[1][Theorem]{%
  \par\vspace{1.5ex}\noindent%
  \textbf{\emph{#1.}} \itshape%
}{\par\vspace{1.5ex}}

\newenvironment{customproof}{%
  \par\vspace{1ex}\noindent%
  \emph{Proof.} \normalfont%
}{\hfill$\square$\par\vspace{1.5ex}}

\begin{document}

\title{Learning Volterra Memory Kernels for Non-Markovian Qubit Dynamics}
\author{Jimmie Adriazola}
\affiliation{School of Mathematical and Statistical Sciences, Arizona State University, Tempe, USA}
\affiliation{Simon A. Levin Mathematical, Computational and Modeling Sciences Center, Arizona State University, Tempe, USA}
\author{Katarzyna Roszak}
\affiliation{FZU - Institute of Physics of the Czech Academy of Sciences, 182 00 Prague, Czech Republic}

\begin{abstract}

 We develop a data-driven framework for identifying non-Markovian equations of motion for open quantum systems, demonstrated here for qubit-environment dynamics.  Starting from the Nakajima–Zwanzig formalism, we vectorize the reduced density matrix into a four-dimensional state vector and cast the dynamics as a Volterra integro-differential equation with an operator-valued memory kernel. The learning task is then formulated as a constrained optimization problem over the admissible operator space, where correlation functions are approximated by rational functions using Padé approximants. We establish well-posedness of the learning problem, ensuring existence of minimizers. To assess performance, we construct synthetic data sets from representative test problems of increasing complexity: (i) exactly solvable pure dephasing, with correlation functions expressed in terms of special functions, (ii) a damped Jaynes–Cummings model with an analytic coherence kernel, (iii) a transverse Born model with frequency-resolved bath integrals and population–coherence coupling, and (iv) a non-rotating-wave quantum Rabi model whose memory kernel has no closed form.  Numerical experiments demonstrate that Padé captures nontrivial temporal structures such as oscillatory memory, algebraic tails, and phase-sensitive coherence transfer, and that the learned models generalize across ensembles of physically admissible initial states.  We perform a parametrization-invariant sensitivity analysis and show that the trajectories are insensitive to the unrecoverable parts of the kernel, so the learned models stay predictive despite severe ill-conditioning in kernel recovery. These  results  together  illustrate that data-driven rational approximation provides an effective route to identifying non-Markovian kernels of practical relevance in quantum technologies.
\end{abstract}

\date{\today}
\maketitle

\section{Introduction}
\label{sec1}
The rapid development of quantum technologies has renewed attention to the accurate modeling of open quantum systems, where a system of interest interacts with an uncontrolled environment. Quantum bits (qubits), such as quantum dots, or molecular excitons all suffer from decoherence and dissipation arising from these couplings, limiting performance in quantum information processing, nanoscale devices, and spectroscopy. The standard Markovian approximation, which leads to Lindblad or Gorini–Kossakowski–Sudarshan–Lindblad (GKSL) master equations, is often inadequate in regimes where system–bath correlations persist on time scales comparable to the system dynamics \cite{BreuerPetruccione2002,deVegaAlonso2017,RivasHuelga2012}. Capturing such \emph{non-Markovian} features remains a central challenge for both theory and simulation.

A principled starting point is the Nakajima–Zwanzig projection operator formalism \cite{Nakajima1958,Zwanzig1960}, which yields a Volterra integro–differential equation for the reduced density matrix with an operator–valued memory kernel. Analytically, the kernel can be expressed in terms of bath correlation functions, but these are rarely tractable beyond toy models or Gaussian environments. This has led to substantial efforts across multiple communities to \emph{learn} memory kernels from data. In molecular dynamics and statistical mechanics, generalized Langevin equations (GLEs) with memory kernels have been identified from trajectory data using projection techniques, kernel regression, and rational function approximations \cite{Mori1965,Zwanzig2001,Chorin2013,LiEtAl2017,JungEtAl2017}. Machine learning methods, ranging from Gaussian processes to deep neural networks, have also been applied to infer nonlocal kernels in coarse–grained models of polymers and biomolecules \cite{WangEtAl2020,NoeClementi2020}. These works demonstrate the feasibility of extracting Volterra–type dynamics directly from time–series data.

Recent attention has turned to learning Markovian and non-Markovian master equations directly. Approaches include fitting time–nonlocal kernels from process tomography \cite{PollockModi2018}, reconstructing transfer tensors \cite{CerrilloCao2014}, and more recently applying machine learning to Lindblad generators with state–dependent corrections \cite{KrastanovJiang2021} and within the so-called GENERIC formalism~\cite{SentzEtAl2025}. These methods highlight both the promise and the difficulty: one must balance expressiveness of the learned kernel with physical constraints such as complete positivity, trace preservation, and Hermiticity. 

Our contribution in this work is indebted to the methods developed by these communities. Here, we synthesize various tools to develop a bare bones,  data–driven framework that uses rational approximants (Padé) to represent bath correlation functions with a forward evaluation of the dynamics handled by a non-local Crank-Nicolson method. Crucially, via a Tikhonov regularization, we ensure stable Volterra kernels while retaining flexibility to capture the dynamic features of open quantum systems. 

Beyond methodology, the physical \emph{relevance} of accurate kernel identification is underscored by several model problems. First, exactly solvable toy models such as a two–level system coupled to a bosonic bath yield correlation functions expressible in terms of special functions \cite{Leggett1987,Weiss1999}, providing valuable benchmarks for algorithms. Second,  a qubit coupled to a single damped cavity mode is exactly solvable in closed form, furnishing an analytic memory kernel against which a learned kernel can be checked directly \cite{JaynesCummings1963,Garraway1997}. Third,  semiconductor quantum dots coupled to phonon baths are experimentally realized testbeds where frequency–resolved bath integrals determine decoherence times \cite{Krummheuer2002,Ramsay2010}, making them central to solid–state quantum technologies.  Fourth, restoring the counter-rotating terms of the cavity coupling produces a model with no closed-form kernel at all, the regime data-driven methods are ultimately built for \cite{ClosBreuer2012,FlindtNovotny2004}.  By designing synthetic data around these representative scenarios, we demonstrate that Padé parameterizations can faithfully recover nontrivial temporal structures characteristic of open qubit dynamics. 

This paper establishes a direct bridge between applied mathematics and quantum engineering by formulating the identification of non-Markovian kernels as a constrained optimization problem over operator spaces. Our analysis establishes existence of minimizers, connects to classical Volterra kernel learning, and validates numerical performance on physically relevant testbeds.  We perform a parametrization-invariant sensitivity analysis to account for the model's predictive power even where kernel recovery is ill-conditioned.  In doing so, we provide both a mathematical framework and practical tools for quantum technologies where non-Markovianity cannot be ignored. Due to the simplicity of our approach, we aim to reveal as clearly as possible the numerical challenges ahead with learning matrix-valued Volterra kernels in open quantum settings.

 The paper is organized as follows. Sec.~\ref{sec2} describes the methodology. Secs.~\ref{section:Problem1}, \ref{section:JC}, \ref{section:Problem2}, and \ref{section:Rabi} apply it to four physical problems of consecutively growing complexity. Sec.~\ref{sec:ident} analyzes the identifiability of the learned kernel, and Sec.~\ref{sec:conc} concludes. 
\section{Overall Methodology\label{sec2}}

The theoretical study of open quantum systems is complicated by the enormous dimensionality of the full system–environment Hilbert space.  
In practice, one is rarely interested in the environment degrees of freedom explicitly, but rather in the properties of the subsystem of interest.  
These can be described by the reduced density matrix
\[
\rho(t) \;=\; \mathrm{Tr}_{E}\,\sigma(t),
\]
obtained by tracing out the environment from the total system–environment density operator $\sigma(t)$.  
This construction dramatically reduces the effective state space, yet retains all the physically relevant information: the reduced density matrix yields the correct statistics for any observable acting on the subsystem alone, rendering the full $\sigma(t)$ unnecessary for the prediction of measurable outcomes \cite{BreuerPetruccione2002,RivasHuelga2012,deVegaAlonso2017}.  
Consequently, the analysis of open quantum dynamics is typically carried out at the level of $\rho(t)$ rather than the joint system–environment state.

A fundamental difficulty is that the evolution of $\rho(t)$ is no longer unitary, since environmental degrees of freedom act as an effective reservoir of dissipation and noise.  
In special situations, the reduced dynamics can be solved exactly.  
One canonical example is the pure dephasing variant spin–boson Hamiltonian, which describes a two-level system linearly coupled to a bath of harmonic oscillators.  
This model serves as a testbed for decoherence, dissipation, and quantum phase transitions, and admits closed-form solutions for certain parameter regimes \cite{Leggett1987,Weiss1999}.  
However, such analytically tractable cases are rare.  

In more general settings, perturbative methods are applied.  
The most common is the Born approximation, which assumes weak system–environment coupling
(and is equivalent to second order perturbative expansion with respect to the coupling) and leads to master equations of Gorini–Kossakowski–Sudarshan–Lindblad (GKSL) type under an additional Markov approximation \cite{Gorini1976,Lindblad1976,BreuerPetruccione2002}.  
These models have been commonly used in many scenarios, including quantum optics
and solid state qubits, but they fail in regimes where system–bath correlations persist on timescales comparable to system evolution.  
Capturing such features requires either going beyond the Markov approximation or 
taking into account higher order processes, and motivates the development of alternative frameworks, such as projection operator techniques \cite{Nakajima1958,Zwanzig1960,Shibata1977,breuer99,breuer01,breuer06}, transfer tensor methods \cite{CerrilloCao2014}, and data-driven approaches.

For example, in the paradigmatic spin–boson model mentioned above, the second-order
approximation is not sufficient to capture the full evolution of the spin qubit in many parameter ranges
\cite{lampo17}. In order to fully capture the evolution of a qubit undergoing pure dephasing due to an interaction with a
bosonic environment, the perturbative approach cannot be truncated nor can the Markov approximation be made. The non-approximate
Nakajima-Zwanzig equation \cite{BreuerPetruccione2002} for
the reduced qubit density matrix $\rho_S(t)$ for this problem is a non-local time equation of the form
\begin{align}
i\frac{d \rho_S(t)}{dt}
   &= [H_S, \rho_S(t)] \notag\\
   &\quad - i \int_0^t 
      \Big([\sigma_z,\, \sigma_z(\tau-t)\rho_S(\tau)]\,C(t-\tau)\Big)\,d\tau .
\end{align}
Here $H_S$ is the Hamiltonian of the system, $\sigma_z$ is the appropriate Pauli matrix, and $C(t-\tau)$ denotes the bath correlation function associated with the oscillator environment.  
This structure illustrates the essential feature: the subsystem dynamics acquire memory terms that integrate over the entire history of the qubit–bath interaction.

More generally, the Nakajima–Zwanzig equation for a qubit can be written in Volterra integro–differential form \cite{Nakajima1958,Zwanzig1960,Shibata1977}:
\begin{equation}\label{eq:VID}
\frac{d\rho}{dt}
   = \mathcal{L}\rho
     + \int_0^t \mathcal{K}(t-\tau)\,\rho(\tau)\,d\tau,
\end{equation}
where $\mathcal{L}$ is the local Liouvillian generating local unitary dynamics, and $\mathcal{K}$ is the memory kernel encoding the influence of the environment.  
The analytic structure of $\mathcal{K}$ is determined by bath correlation functions, but explicit expressions are typically unavailable beyond Gaussian or perturbative settings \cite{deVegaAlonso2017,RivasHuelga2012}.
Note, that for the form of eq.~(\ref{eq:VID}) to be sufficient to describe a physical scenario, two assumptions
have to hold \cite{BreuerPetruccione2002}. One is that odd moments of the interaction Hamiltonian 
are zero when taken with respect to the initial state of the environment, and the other that the initial
system-environment state has product form. Both assumptions are commonly fulfilled. 

To recast eq.~\eqref{eq:VID} in a form more amenable to numerical discretization and data–driven inference, we vectorize the $2\times 2$ qubit density matrix $\rho$ into a four–dimensional state vector
\[
\mathbf{x}(t) \;=\;
\big[\rho_{00}(t),\;\rho_{11}(t),\;\rho_{01}(t),\;\rho_{10}(t)\big]^{\intercal}.
\]
In these coordinates the dynamics reduce to a linear Volterra system
\begin{equation}\label{eq:State}
\frac{d\mathbf{x}}{dt}
   = A \mathbf{x}(t)
     + \int_0^t B(t-\tau)\,\mathbf{x}(\tau)\,d\tau,
\end{equation}
with $A\in\mathbb{C}^{4\times 4}$ representing the instantaneous generator and $B(\cdot)$ a matrix–valued kernel encoding time–nonlocal correlations.  
Equation~\eqref{eq:State} is the basic form we adopt for our learning problem.

The central objective of this work is to identify $A$ and $B$ directly from data.  
For $A$, the hypothesis class is straightforward: it is a fixed $4\times 4$ complex matrix, requiring the determination of 16 parameters.  
For $B$, the hypothesis class is substantially richer: each entry $B_{ij}(t)$ is a time–dependent correlation function whose structure must be captured in a way that is both numerically tractable and physically consistent.

In the examples following this section, it will become clear that a symbolic library of elementary transcendental functions will not suffice to capture the behavior of the correlation kernel. At the same time, we do not seek to overparametrize the problem or use difficult to interpret architectures such as deep neural networks. In this work, we propose to use Padé approximants to model the behavior of each correlation function.  A Pad\'e approximant $[q/r]$ of a function is the ratio of a degree-$q$ polynomial to a degree-$r$ polynomial, with coefficients chosen to match the
function. More explicitly, we model each entry of the correlation function by the $[q/r]$ Pad\'e approximant
\begin{equation}\label{eq:PadeHyp}
B_{i,j}^{[q/r]}(t;\xi):=\frac{\sum_{k=0}^{q}\xi_k t^k}
{\sum_{k=q+1}^{q+r+1}\xi_k t^{k-q-1}},
\end{equation}
where the explicit parametrization is the vector $\xi\in\mathbb{C}^{q+r+2}$. 
  Assuming each entry in the correlation kernel has the same order $[q/r]$, the parametrization tensor of the correlation kernel is given by $\Xi\in\mathbb{C}^{4\times 4\times (q+r+2)}.$ Together with the 32 real parameters needed to learn $A$, the total number of real parameters is $N_{\xi}=32(q+r+2)+32$.

Before moving forward, we remark that the use of a Padé approximant at this stage is treated as a computational choice whose viability will be demonstrated throughout this paper.  Briefly, our rationale for pursuring this ansatz is simple. The numerator sets the short-lag behavior and the denominator controls the decay, so a low-order rational captures both oscillatory and
algebraically decaying kernels that neither a polynomial nor a sum of exponentials could. We perform quantitative tests against these other ansatzes in Appendix~\ref{app:details}.

With these hypotheses in hand, we define the admissible operator space that we propose to search over. First, recall that the Sobolev space, $H^s([0, T])$, is a Hilbert space defined as
\[
H^s([0, T]) := \left\{ f \in L^2([0, T]) \ :\ \frac{d^k f}{dt^k} \in L^2([0, T]) \right\},
\]
with corresponding norm
$$
\|f\|_{H^s([0, T])}^2 := \sum_{k=0}^s \left\| \frac{d^k f}{dt^k} \right\|_{L^2([0, T])}^2,
$$
for all integers $0 \leq k \leq s$. Then, the proposed operator hypothesis space for our data-driven learning problem is given in compact form as
\begin{equation}\label{eq:OperatorSet}
\mathcal{O} = \left\{
(A, B) \ :\
A \in \mathbb{C}^{4 \times 4}, \
B \in H^1\big([0, T]; \mathbb{C}^{4 \times 4}\big)
\right\}
\end{equation}
where the Bochner space
$$
H^s\left(\Omega ; \mathbb{C}^{4 \times 4}\right):=\left\{B: \Omega \rightarrow \mathbb{C}^{4 \times 4} \mid B_{i j} \in H^s(\Omega) \ \forall \ i, j\right\}
$$
is a Sobolev space of functions, defined on the Borel measurable set $\Omega$, taking values in $\mathbb{C}^{4 \times 4}$. No further structure on the matrices $A$ and $B$ is assumed at this level, but will be enforced and discussed in the context of the example problems throughout this paper.

We now define the objective functional that models the data-driven discovery of Nakajima-Zwanzig equations. Our first contribution to the objective functional is the loss function
\begin{equation}
    \mathcal{J}_{\rm loss}[A,B]=\sum_{j}\int_0^T\left|x_j^{\rm data}(t)-x_j^{\rm learned}(t)\right|^2dt
\end{equation}
where $\mathbf{x}^{\rm data}$ is a time-series of data, either synthetic or experimental, while $\mathbf{x}^{\rm learned}$ is the learned dynamics satisfying eq.~\eqref{eq:State} corresponding to the search over the operator space $\mathcal{O}.$  

The second contribution to the objective functional is a weighted Tikhonov regularization, balancing an $L^2([0,T])$ penalization  on the function  and an $L^2([0,T])$ penalization on the time derivatives. A regularization is necessary given the ill-posed nature of learning kernels that satisfy a Volterra equation of the first kind. The regularization we used is expressed as
\begin{equation}\label{eq:reg}
    \mathcal{J}_{\rm reg}[B]=\sum_{i,j}(1-\beta)\left\|B_{i,j}\right\|_{L^2([0,T])}^2+\beta\left\|\frac{dB_{i,j}}{dt}\right\|_{L^2([0,T])}^2.
\end{equation}
We note that, theoretically, for the problem to remain well-posed in the space $\mathcal{O}$, it is sufficient to penalize with the $H^1\left([0,T];\mathbb{C}^{4\times4}\right)$ norm; see Appendix~\ref{section:WP}. However, to obtain desirable numerical results, we found it advantageous to weigh these two norms on $L^2([0,T])$ with an empirically chosen weighting parameter $\beta\in(0,1).$

Thus, with a hypothesis space and an objective functional, we can now pose the learning of non-Markovian open quantum systems modeled by the Nakajima-Zwanzig equation~\eqref{eq:VID} as the constrained optimization problem
\begin{equation}\label{eq:RegProb}
\min_{\{A,B\}\in\mathcal{O}} \mathcal{J}[A,B]=\min_{\{A,B\}\in\mathcal{O}}(1-\alpha)\mathcal{J}_{\rm loss}[A,B]+\alpha\mathcal{J_{\rm reg}}[B]
\end{equation}
subject to eq.~\eqref{eq:State}, where $\alpha\in(0,1)$ is a parameter that balances between operator discovery and smoothness of the discovered operators. 

 The parametrization turns the search over the operator space $\mathcal{O}$ into a finite-dimensional optimization over the real vector $\xi\in\mathbb{R}^{N_{\xi}}$. Because the objective is formulated in terms of a 
least-squares, we minimize it with the Levenberg--Marquardt method
through MATLAB's \texttt{lsqnonlin}. For the scalar dephasing fit of Sec.~\ref{section:Problem1}, the problem is well conditioned so an unconstrained quasi-Newton step via \texttt{fminunc}
suffices.  We discuss the computational choices made in further designing and implementing this abstract regression problem in the following examples.

\section{Test Problem 1: Spin-boson pure dephasing\label{section:Problem1}}

Pure dephasing (or pure decoherence) \cite{zurek03,roszak15,roszak18} occurs when the interaction between a system and an environment does not involve 
the exchange of energy, but is related to the transfer of information about the system state to the environment
\cite{zurek03}. This means that the environment cannot affect the occupations of the system in the so called
pointer basis (the basis that diagonalizes both the system Hamiltonian and the interaction Hamiltonian),
while it causes a decay of the off-diagonal elements of the density matrix in this basis. If the system is 
a qubit, then the environment-induced evolution is limited to a single element of the density matrix
(the coherence). Thus pure dephasing of the qubit is described by the simplest version of Volterra integro-differential equation, eq.~(\ref{eq:VID}).

The spin-boson model \cite{Leggett1987,guo12,cai14,ferialdi17} is one of the canonical examples of
a model that leads to qubit pure dephasing. The Hamiltonian is given by 
\begin{equation}
    \label{ham1}
    H=\frac{\varepsilon}{2}\sigma_z
    +\sum_{k}\omega_kb_k^{\dagger}b_k
    +\sigma_z\sum_{k}\left(f_kb_k^{\dagger}+f_k^*b_k\right).
\end{equation}
Here, the first term describes the free evolution of the qubit (spin), the second term is the free Hamiltonian
of the bosonic environment, and the third term describes the interaction. The operators $b_k^{\dagger}$
and $b_k$ are bosonic creation and annihilation operators corresponding to wave-vector $k$, respectively,
and the corresponding energies are given by $\omega_k$. In the interaction, $f_k$ denote the coupling 
constants, while 
$\varepsilon$ is the energy difference between the two spin states.

This model is exactly solvable \cite{BreuerPetruccione2002} and for the environment initially 
in a Gibbs state corresponding to temperature $T$, the interaction-picture evolution of the qubit coherence 
is given by an equation of the form
\begin{equation}\label{eq:Test1State}
\frac{d \rho_{01}(t)}{d t}=-\int_0^t \rho_{01}(s)C(t-s)ds.
\end{equation}
Here, the correlation function is given by the following integral
\begin{equation}\label{eq:Test1C}
C(t-s)=\int_0^{\infty} (2 n_{\rm B}(\omega)+1)J(\omega)\cos [\omega(t-s)]d\omega,
\end{equation}
where the Bose-Einstein occupation number and the spectral density are given by 
\begin{equation}
n_{\rm B}(\omega)=\left(e^{\frac{\omega}{k_{\rm B}T}}-1\right)^{-1},\quad J(\omega)=2\sum_{k}|f_k|^2\delta(\omega-\omega_k),    
\end{equation}
respectively. Here $k_{\rm B}$ is the Boltzmann constant.
In the following we assume that the qubit is initially in an equal superposition of the pointer states.

The specific form of the coupling constants, and consequently the specific form of the spectral density, depends
on the physical problem under study. We follow Ref.~\cite{lampo17} and take
\begin{equation}
\label{jotodomega}
    J(\omega)=g\frac{\omega^p}{\Lambda^{p-1}}\exp \left(-\frac{\omega}{\Lambda}\right),
\end{equation}
with the cut-off frequency $\Lambda =1$. $p$ denotes the Ohmicity parameter; for $p\in(0,1)$ we are dealing with
a sub-Ohmic environment, for $p=1$ the environment is Ohmic, and for $p>1$ it is super-Ohmic. The features
that can be displayed by the evolution of the qubit qualitatively depend on the value of this parameter.
The parameter $g$ is responsible for the overall strength of the coupling and in this section
is taken $g=1$.

To compute the correlation function, we are tasked with computing the integral over $\omega$ exactly. 
To this end, denote $k(\omega,t)=J(\omega)\cos(\omega t)$ so that we can rewrite the correlation function as
\begin{align}
C(t)=2\int_0^\infty k(\omega,t)d\omega&+\int_0^\infty k(\omega,t)n_{\rm B}(\omega)d\omega\nonumber\\
:=I_1(t)&+I_2(t).\nonumber
\end{align}
With assumed $k_{\rm B}T=1$, the first integral is found to be 
$$
I_1(t)=\Gamma (p+1) \left(t^2+1\right)^{-\frac{p}{2}-\frac{1}{2}} \cos \left((p+1) \tan ^{-1}(t)\right),
$$
while the second integral can be handled using special functions by observing that
\begin{align}
I_2(t)&=2\int_0^{\infty}\frac{e^{-\omega}\omega^p\cos(\omega t)}{e^{\omega}-1}\,d\omega\nonumber \\
&=2\Re\left(\int_0^{\infty}\frac{\omega^pe^{-(2-it)\omega}}{1-e^{-\omega}}\,d\omega\right)\nonumber \\
&=2\Re\left(\Gamma(p+1)\zeta(p+1,2-it)\right),\phantom{\int}\nonumber
\end{align}
where $\Gamma$ is the standard gamma function, and $\zeta$ is the Hurwitz zeta function. We see clearly that, for this learning problem, regression on the space of polynomials is likely to be insufficient for reproducing the behavior of these special functions. For this reason, we introduce a Padé approximant, given by Equation~\eqref{eq:PadeHyp}. Specifically, we use a [4/4] Padé approximant and, consequently, try to learn ten real parameters.

We prepare synthetic data on a time domain of $t\in[10^{-6},3]$ with 64 grid points. Using an initial time of $10^{-6}$ is done to avoid the removable singularity of Equation~\eqref{eq:Test1C} at $t=0.$ To numerically construct the correlation function~\eqref{eq:Test1C}, we use MATLAB's built-in Hurwitz zeta function by evaluating \texttt{hurwitzZeta(p+1, 2-1i*t)}, along with evaluating the remaining special and elementary functions in standard ways. We then numerically integrate Equation~\eqref{eq:Test1State}, from an initial condition of $\rho_{01}(0)=\frac{1}{2},$ by using the non-local Crank-Nicolson scheme outlined in Appendix~\ref{section:NM}. 

\begin{figure}[!tb]
\begin{centering}
\subfigure{\includegraphics[width=0.45\textwidth]{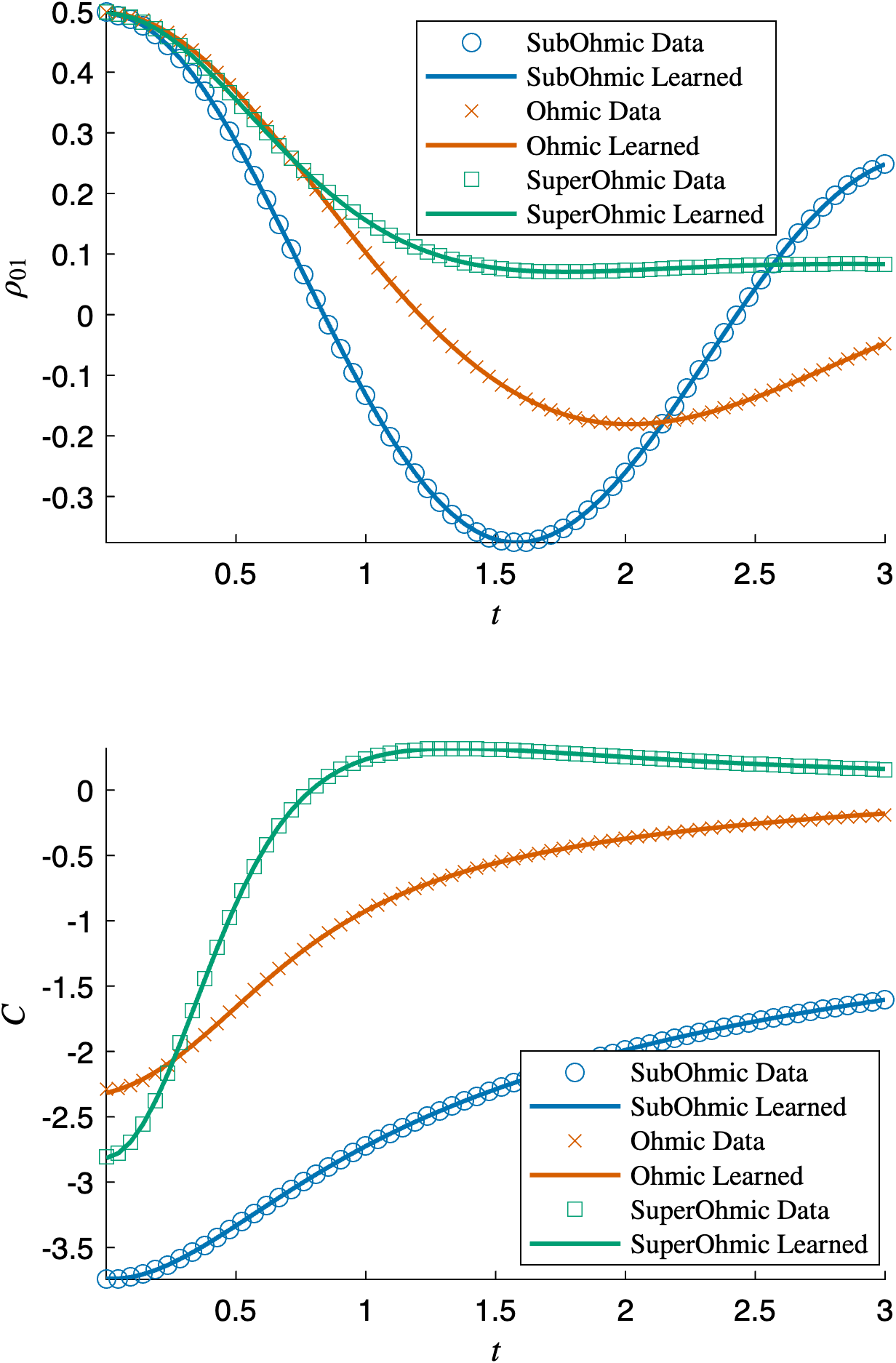}}
\end{centering}
\caption{A numerical solution of Problem~\eqref{eq:RegProb}, with $\alpha=0$ and constrained by Equation~\eqref{eq:Test1State}, using the [4/4] Padé approximant expressed by Equation~\eqref{eq:PadeHyp}. We use a subohmic parameter of $p=1/2$ and a superohmic parameter of $p=2.$ The construction of the synthetic data used in this study is discussed in the main text.}\label{fig:Test1}
\end{figure}

For this problem, we observe empirically through numerical experimentation that a regularization on the correlation function is unnecessary. Therefore, we solve the unregularized optimization problem~\eqref{eq:RegProb}, that is, with 
$\alpha=0$,
and for each of the three different physical regimes governed by the parameter Ohmic spectral parameter $p$. We study the subohmic $0<p<1,$ ohmic ($p=1$), and superohmic regimes $(p>1)$, and in all examples, accurately reproduce the dynamics furnished by $C(t)$ using the optimal Padé approximant $P(t,\xi^*).$ Our results are visualized by Figure~\ref{fig:Test1}.

Indeed, we observe a typical loss, that is $\mathcal{J}_{\rm reg}[\xi_*]$, on the order of $10^{-9}$. Moreover, the realization of the dynamics, by virtue of the fast evaluations of Padé approximants, is on the order of hundredths of a second on an average Macbook Air laptop. Meanwhile, the synthetic data took about 6 seconds to prepare. Furthermore, the reconstruction of the correlation function in all three parametric regimes is accurate, despite only being trained on one trajectory, therefore, we do not investigate how our trained model generalizes given the uniqueness of integral curves from the dynamics.

Now, we investigate the effect of corrupting our synthetic data with noise. We perturb the Ohmic data, which can be seen in the top panel of Figure~\ref{fig:Test1} with a 10\% relative amplitude, that is, $\rho^w_{01}(t)=(1+w(t))\rho^{\rm data}_{01}(t)$ where $w(t)$ is uniformly sampled on $[-1/10,1/10]$ at each time $t$. The trained dynamics, without regularization, typically yields a loss that evaluates on the order of $10^{-4}.$ 

\begin{figure}[!htb]
\begin{centering}
\subfigure{\includegraphics[width=0.45\textwidth]{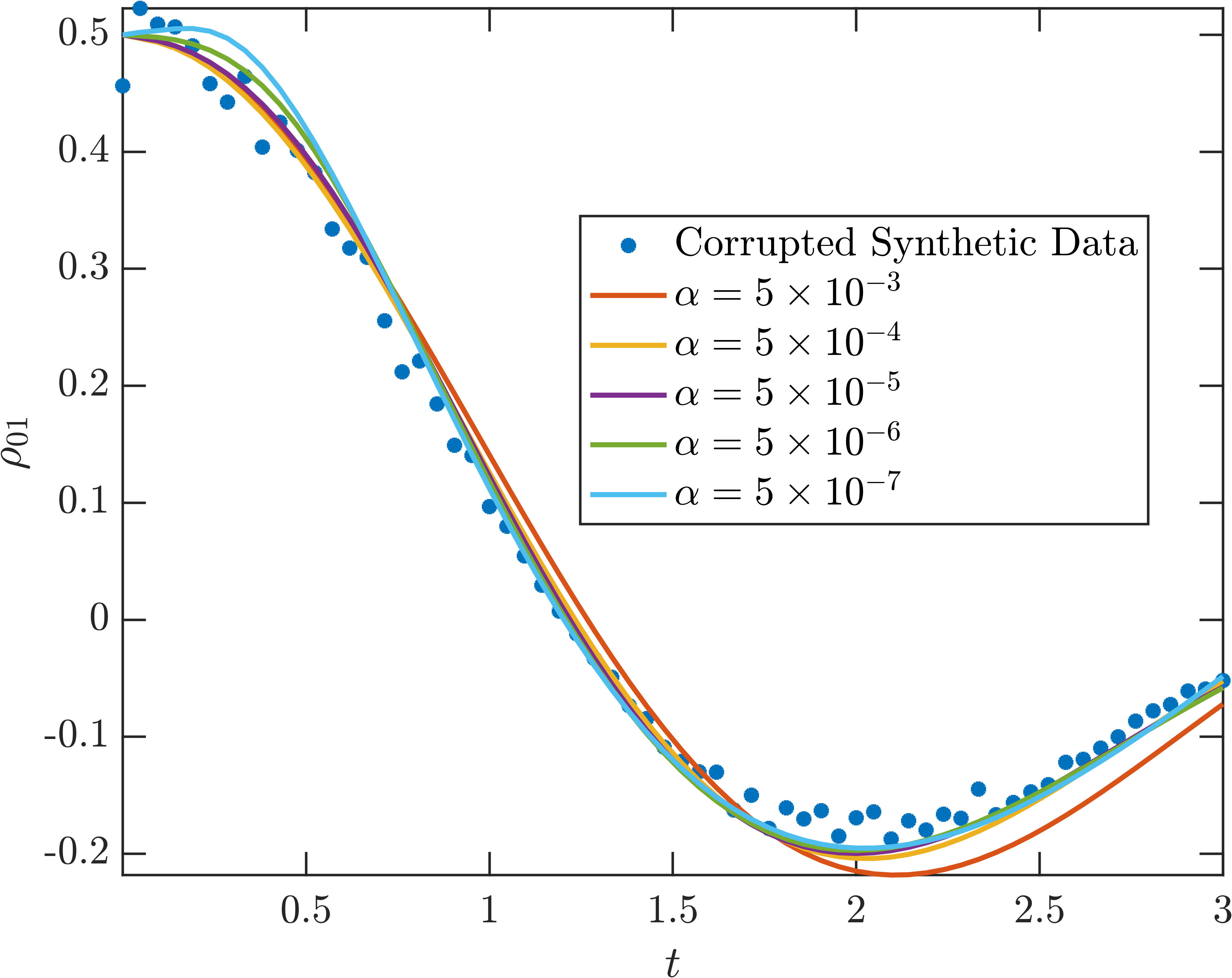}}
\subfigure{\includegraphics[width=0.45\textwidth]{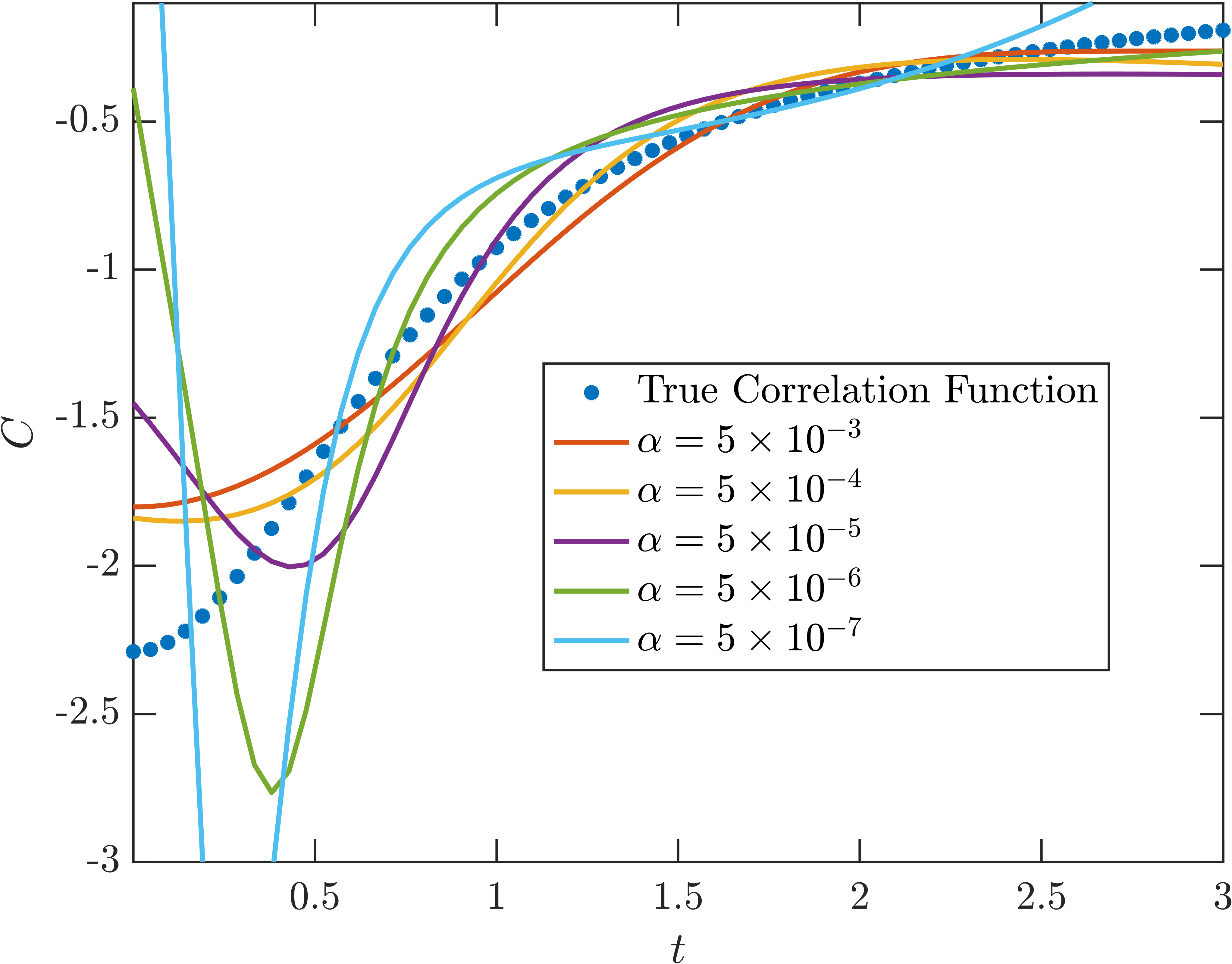}}
\end{centering}
\caption{We display the effect of noise on the regression by  performing the same study as in the Ohmic case shown in Figure~\ref{fig:Test1}, yet introduce corruption in the synthetic data in the way described in the main text. We observe that even a small Tikhonov regularization  does well to smooth out singular features of the correlation function, while maintaining a fairly reasonable fit to the synthetic data. Decreasing the regularization leads to a monotonically decreasing singular feature in the bottom panel near $t=0.5$.}\label{fig:Noise}
\end{figure}

Despite this somewhat small loss, we observe that the correlation function begins to exhibit a fast oscillation near $t=0.5$. To regularize this behavior, we use a homogeneous Sobolev regularization ($\beta=1$ in Equation~\eqref{eq:reg}) while varying the weighting parameter $\alpha$ in eq.~\eqref{eq:RegProb}. As can be seen in Figure~\ref{fig:Noise}, the regularization does well to smooth out this fast oscillation, yet maintains an evaluation of the loss function that remains within the same order of magnitude as the unregularized case.

\section{Test Problem 2: A qubit and a damped cavity mode (Jaynes--Cummings)}\label{section:JC}

We now consider a two-level system coupled to a single quantized cavity mode under the
rotating-wave approximation~\cite{scully97,meystre07}, with the cavity itself leaking into a flat continuum~\cite{JaynesCummings1963}.
Setting $\hbar=1$, the qubit--mode Hamiltonian is~\cite{JaynesCummings1963}
\begin{equation}\label{eq:JC-ham}
H=\frac{\omega_0}{2}\sigma_z+\omega_c\,a^\dagger a
 +g\big(\sigma_+ a+\sigma_- a^\dagger\big),
\end{equation}
where $\sigma_\pm$ are the qubit raising and lowering operators, $a,a^\dagger$ are the
mode ladder operators with $[a,a^\dagger]=1$ (identity on the mode space), $\omega_0$ is the qubit splitting,
$\omega_c$ the mode frequency, and $g$ the dipole coupling. 

The cavity is lossy, and this amounts to a mode damping at linewidth $\kappa$ by a
Lindblad dissipator, so the joint qubit--mode state $\chi$, a density operator on
$\mathbb{C}^2\otimes\mathcal{F}$ with $\mathcal{F}$ the mode Fock space, obeys
\begin{equation}\label{eq:JC-lindblad}
\begin{aligned}
\dot\chi&=-i[H,\chi]+\kappa\,\mathcal{D}[a]\chi,\\
\mathcal{D}[a]\chi&=a\chi a^\dagger-\tfrac{1}{2}\big\{a^\dagger a,\chi\big\},
\end{aligned}
\end{equation}
where $\{X,Y\}=XY+YX$ denotes the anticommutator. The qubit is recovered by tracing the mode out of the joint state,
$\rho(t)=\mathrm{Tr}_{\mathcal{F}}\,\chi(t)=\sum_{n}\langle n|\chi(t)|n\rangle$,
the partial trace over the Fock basis $\{|n\rangle\}$~\cite{BreuerPetruccione2002}.

A single Lindblad-damped mode reproduces the exact reduced dynamics of a qubit
coupled to a continuum reservoir of Lorentzian spectral density
\begin{equation}\label{eq:JC-J}
J(\omega)=\frac{1}{2\pi}\,\frac{\gamma_0\kappa^2}{(\omega-\omega_c)^2+\kappa^2},
\end{equation}
centered at the mode frequency $\omega_c$, with $\gamma_0$ the Markovian-limit
decay rate of the qubit and the cavity linewidth $\kappa$ setting the reservoir
width. This is an example of what is referred to as the pseudomode correspondence in the literature~\cite{Garraway1997,ferialdi17}. The practical implication is that the
Markovian damping of one discrete mode reappears as a finite-width spectral
feature for the qubit, and that finite width
is what renders the reduced dynamics non-Markovian. 

Precisely, the reservoir correlation function is the single exponential
\begin{equation}\label{eq:JC-f}
f(\tau)=\frac{\gamma_0\kappa}{2}\,e^{-(\kappa-i\delta)\tau},
\qquad \delta=\omega_0-\omega_c .
\end{equation}
Restricting to the single-excitation sector, with the reservoir in vacuum and at
most one quantum shared between qubit and mode, the excited-state amplitude $G(t)$
obeys the convolution equation
\begin{equation}\label{eq:JC-amp}
\dot G(t)=-\int_0^t f(t-\tau)\,G(\tau)\,d\tau, \qquad G(0)=1,
\end{equation}
whose exponential kernel gives the closed form
\begin{equation}\label{eq:JC-G}
G(t)=e^{-at/2}\!\left[\cosh\frac{dt}{2}+\frac{a}{d}\sinh\frac{dt}{2}\right],
\end{equation}
with $a=\kappa-i\delta$ and $d=\sqrt{a^2-2\gamma_0\kappa}$; the amplitude
equation and its solution are standard, and we follow
Ref.~\cite{BreuerPetruccione2002}. In the strong-coupling regime, $d$ is
imaginary and the excited-state
weight $|G(t)|^2$ revives. This is the signature of non-Markovianity in which excitation
kicks back from the reservoir to the qubit.

Meanwhile, the reduced map is the amplitude-damping channel
\begin{equation}\label{eq:JC-channel}
\begin{aligned}
\rho_{11}(t)&=\rho_{11}(0)\,|G(t)|^2,\\
\rho_{01}(t)&=\rho_{01}(0)\,\overline{G(t)},\\
\rho_{00}(t)&=1-\rho_{11}(t),
\end{aligned}
\end{equation}
written in the vectorization $\mathbf{x}=[\rho_{00},\rho_{11},\rho_{01},\rho_{10}]^\top$,
with index $0$ the ground level and $1$ the excited level. Since $|G(t)|\le1$ at all
times, the channel is completely positive (CP) throughout, so the synthetic training
data are CP by construction with no enforcement on our part.

The model conserves the total excitation number $\sigma_+\sigma_-+a^\dagger a$, so
the reduced generator commutes with the $\sigma_z$ phase rotation and the kernel
\begin{equation}\label{eq:JC-kernel}
B_{\mathrm{JC}}(\tau)=
\begin{pmatrix}
 0 & -b_{22}(\tau) & 0 & 0\\
 0 &  b_{22}(\tau) & 0 & 0\\
 0 & 0 & b_{33}(\tau) & 0\\
 0 & 0 & 0 & \overline{b_{33}(\tau)}
\end{pmatrix}
\end{equation}
does not mix populations with coherences. Moreover, since the coherence is linear, $\rho_{01}=\rho_{01}(0)\,\overline{G}$,
conjugating \eqref{eq:JC-amp} gives a closed convolution for $\rho_{01}$. Indeed, the
coherence kernel inherits the reservoir correlation as
\begin{equation}\label{eq:b33}
b_{33}(\tau)=-\overline{f(\tau)}.
\end{equation}

\begin{figure}[t]
\centering
\includegraphics[width=\columnwidth]{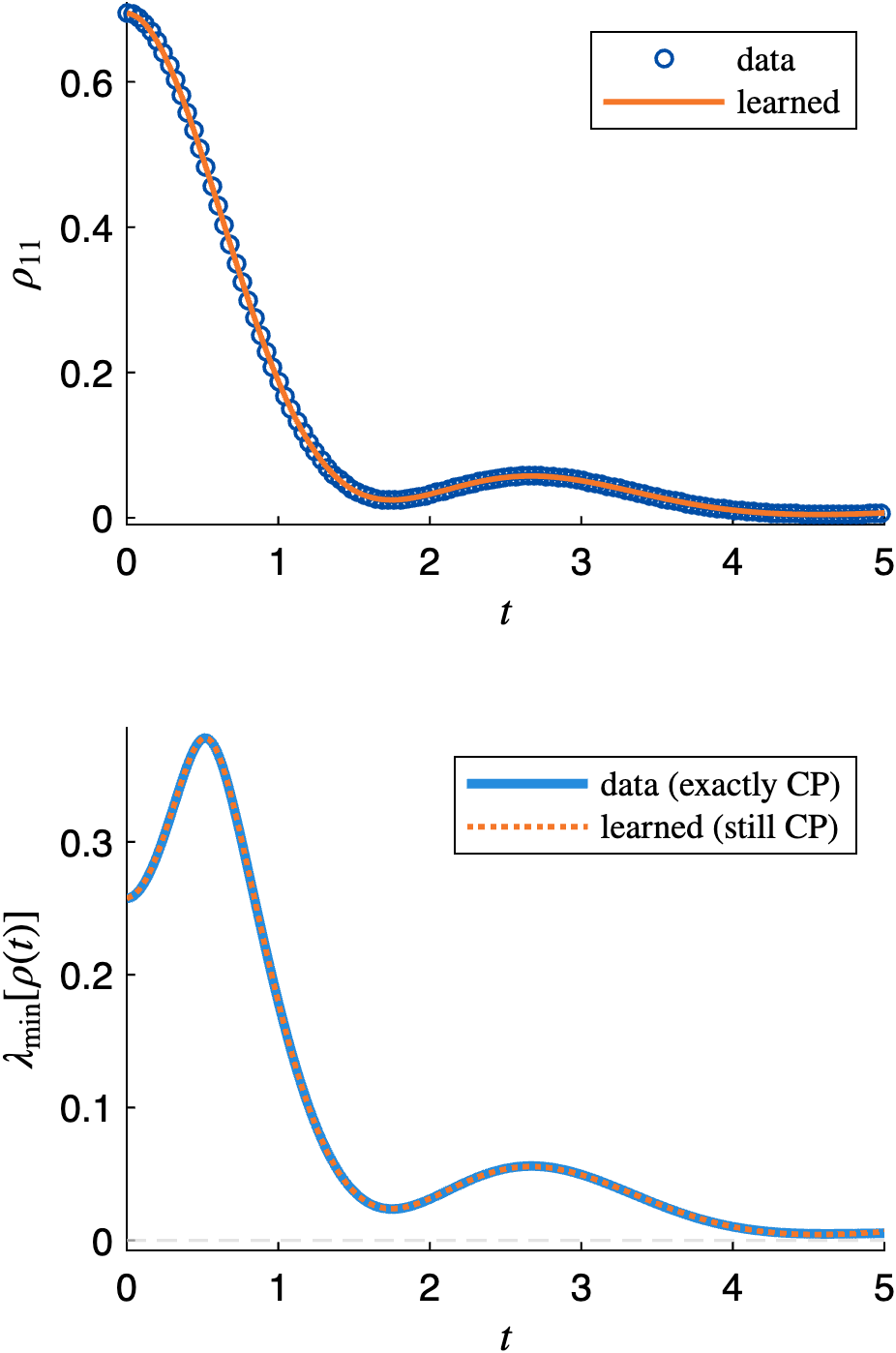}
\caption{Damped Jaynes-Cummings test ($\gamma_0=3$, $\kappa=1$, $\delta=0.5$,
$R=6$). Top: learned model (line) against a held-out trajectory (circles) for the
excited population $\rho_{11}$, including the revival near $t\approx1.75$. Bottom:
minimum eigenvalue $\lambda_{\min}[\rho(t)]$ showing the learned dynamics (dotted) stay on the ground truth (solid) while both remaining physical.}
\label{fig:jc-fit}
\end{figure}

For our numerical experiments, we take $\gamma_0=3$, $\kappa=1$, $\delta=0.5$, giving quality factor
$R=2\gamma_0/\kappa=6$, deep in the strong coupling regime, where we see (in the state response in Figure~\ref{fig:jc-fit}) $|G(t)|^2$ revives near
$t\approx1.75$. Initial states are drawn from the physical set by sampling
\[
R=\begin{pmatrix}
s_1 & i s_2\\
-i s_2 & 1-s_1
\end{pmatrix},
\qquad s_1,s_2\sim \textrm{Uniform}(0,1),
\]
which is an effective projection onto the positive semi-definite cone with unit trace. This matrix state is then vectorized to the initial state vector $x_0\in\mathbb{C}^4.$ We then integrate the channel \eqref{eq:JC-channel} on $t\in[0,5]$ with $250$ points, so the training data are CP by
construction. We model $b_{33}$ by a complex $[3/3]$ Pad\'e and $b_{22}$ by a real
$[3/3]$ Pad\'e (denominator constant unity, $21$ real parameters), minimize the
regularized objective \eqref{eq:RegProb} over $12$ training trajectories. We found that training over multiple trajectories improves generalization by at least an order of magnitude in this matrix-valued kernel setting.

\begin{figure}[t]
\centering
\includegraphics[width=\columnwidth]{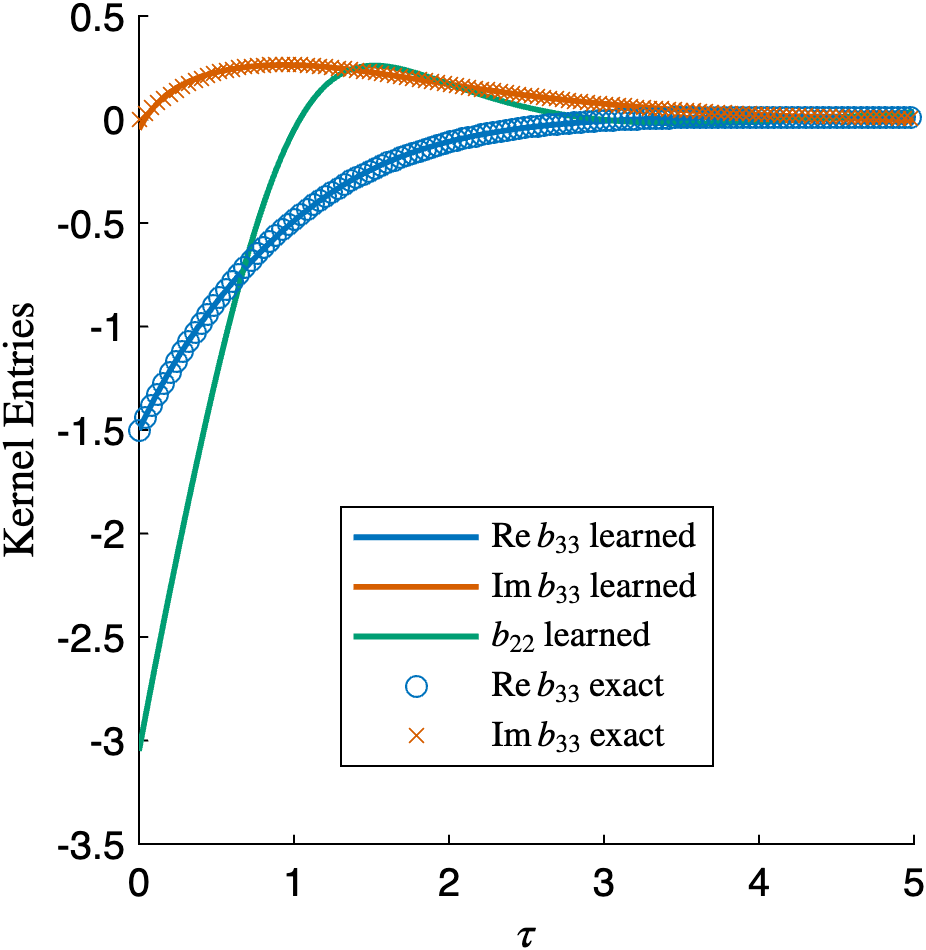}
\caption{Kernel recovery for the damped Jaynes-Cummings test, all entries on one
axis. The learned coherence kernel $b_{33}(\tau)$ (solid) is shown against the
analytic $-\overline{f(\tau)}$ (markers) for both real and imaginary parts,
recovered to a relative error of $1.4\times10^{-2}$ that is roughly uniform across
the window, together with the learned population kernel $b_{22}(\tau)$, which has no
simple closed form. As in Figure~\ref{fig:Test1}, we observe another accurate kernel identification.}
\label{fig:jc-kernel}
\end{figure}

To discuss generalization, we use the empirical risk (or held-out misfit)
\begin{equation}\label{eq:EmpRisk}
\mathcal{R}[A,B]=\mathbb{E}_{\rho(0)\sim \varrho}\mathcal{J}_{\rm loss}[A,B].
\end{equation}
At $\alpha=10^{-6}$, $\beta=0.5$ the method recovers $b_{33}=-\bar f$ to a
relative error of $1.4\times10^{-2}$, roughly uniform across the lag window, and a held-out misfit of $1.8\times10^{-6}$ with
minimum eigenvalue $6.1\times10^{-5}$ across a 40 state ensemble. Additionally, the regularization we used to find this result is essential. Without the penalty ($\alpha=0$), the misfit sits at a fairly large value of $1.4\times10^{-2}$ and the minimum eigenvalue collapses to $-3.5\times10^{5}$. We found that $\alpha>0$ together with the derivative penalty
($\beta>0$) worked best, while too large
an $\alpha$ pulls $b_{33}$ off $-\bar f$ and grazes the cone near $-10^{-3}$, losing the CP structure. Our results are summarized concisely in Table~\ref{tab:jc-reg}.

\begin{table}[t]
\centering
\setlength{\tabcolsep}{6pt}
\renewcommand{\arraystretch}{1.2}
\caption{Regularization trade-off for the damped Jaynes-Cummings fit
($\gamma_0=3$, $\kappa=1$, $\delta=0.5$). Held-out misfit and worst-case
minimum eigenvalue over a $40$-state ensemble.}
\label{tab:jc-reg}
\begin{tabular*}{\columnwidth}{@{\extracolsep{\fill}}|c|c|c|c|c|}
\hline
$\alpha$ & $\beta$ & risk & worst $\lambda_{\min}$ & regime \\
\hline
$0$       & \textemdash & $\sim\!10^{-2}$ & $\hspace{-6pt}\sim\!-10^{5}$  & blow-up \\
$10^{-6}$ & $0.5$       & $\sim\!10^{-6}$ & $\sim\!+10^{-5}$ & recovered, CP \\
$10^{-3}$ & $0.9$       & $\sim\!10^{-3}$ & $\sim\!-10^{-3}$ & over-regularized \\
\hline
\end{tabular*}
\end{table}

\section{Test Problem 3: The transverse Born model}\label{section:Problem2}

A similar qubit-environment Hamiltonian to that of Test Problem 1 can be used to model an interaction that
does involve energy exchange and leads not only to pure dephasing, but involves the evolution of qubit occupations
in any basis \cite{divincenzo05,wu17,gulacsi23}. The condition is that the free qubit Hamiltonian
does not commute with the interaction term \cite{strzalka24}. To fulfill this condition, we 
exchange the interaction in eq.~(\ref{ham1}) as follows,
\begin{equation}
    \label{ham2}
    \sigma_z\sum_{k}\left(f_kb_k^{\dagger}+f_k^*b_k\right)\rightarrow
    \sigma_x\sum_{k}\left(f_kb_k^{\dagger}+f_k^*b_k\right).
\end{equation}
Note, that the change in the Hamiltonian is small (limited to the exchange of the Pauli matrix
that governs the effect of the interaction on the qubit), but it leads to a fundamental change
of the nature of the decoherence.

Thus, now both the occupations and the coherences of the qubit state affected by the interaction
with the environment, and their evolution (in the interaction picture) is governed
by the following equations,
\begin{subequations}
\label{eq:rho-clean}
\begin{align}
\frac{d\rho_{00}}{dt} &= - \int_0^t \!\! ds \int_0^\infty \!\! d\omega \, J(\omega) 
\cos\big((\varepsilon_0{-}\omega)(t{-}s)\big) \notag \\
&\quad \times \left[ \rho_{00}(s)(2n(\omega){+}1) - \rho_{11}(s)(n(\omega){+}1) \right],
\\
\frac{d\rho_{01}}{dt} &= - \frac{1}{2} \int_0^t \!\! ds \int_0^\infty \!\! d\omega \, J(\omega) 
(2n(\omega){+}1) \notag \\
&\quad \times \big[\, \rho_{01}(s) e^{-i(\varepsilon_0{-}\omega)(t{-}s)} \notag \\
&\qquad\quad - \rho_{10}(s) e^{-2i\varepsilon_0 t} e^{i(\varepsilon_0{-}\omega)(t{-}s)} \,\big],\end{align}
\end{subequations}
together with the closure relations $\rho_{00}(t)+\rho_{11}(t)=1$ and $\rho_{10}(t)=\rho_{01}^*(t)$.
Here, it was assumed that the initial state of the 
environment is a thermal equilibrium state with respect to its free Hamiltonian.

The population equations are already convolutions in the lag $t-s$, but the
$\rho_{10}$ term in eq.~\eqref{eq:rho-clean} carries an explicit factor
$e^{-2i\varepsilon_0 t}$ that depends on absolute time, so the coherence block is
two-time and eq.~\eqref{eq:rho-clean} is not yet of the convolution form
\eqref{eq:State}. We remove this factor by passing to the co-rotating coherence
variables
\begin{equation}\label{eq:corot}
\tilde\rho_{01}(t)=e^{i\varepsilon_0 t}\rho_{01}(t),\qquad
\tilde\rho_{10}(t)=e^{-i\varepsilon_0 t}\rho_{10}(t),
\end{equation}
with the populations unchanged. The phases $e^{\pm i\varepsilon_0 t}$ carried by
$\rho_{01}$ and $\rho_{10}$ cancel the $e^{-2i\varepsilon_0 t}$ in the cross term,
and every bracket in eq.~\eqref{eq:rho-clean} reduces to a function of $t-s$ alone.
In the variables $\mathbf{x}=[\rho_{00},\rho_{11},\tilde\rho_{01},\tilde\rho_{10}]^{\top}$
the dynamics are then exactly of the form \eqref{eq:State}, with the frame rotation
carried by a local generator
\begin{equation}\label{eq:corotA}
A=\mathrm{diag}\!\left(0,\,0,\,i\varepsilon_0,\,-i\varepsilon_0\right),
\end{equation}
and a genuine convolution kernel of the block form

\begin{equation}\label{eq:blockB}
B=\left(\begin{array}{cccc}
B_{00} & B_{11} & 0 & 0 \\
-B_{00} & -B_{11} & 0 & 0 \\
0 & 0 & B_{01} & B_{10} \\
0 & 0 & B_{10}^{*} & B_{01}^{*}
\end{array}\right).
\end{equation}

This convolution form,
not available in the original interaction frame, is what lets us learn both $A$ and
$B$ with the lag-only ansatz \eqref{eq:PadeHyp} against the optimization problem
\eqref{eq:RegProb}. We see something similar in Sec.~\ref{section:Rabi}, where the
qubit splitting is again carried by $A$ and the kernel is a pure convolution. This desirable block matrix structure in the correlation kernel is one reason why we vectorize the density matrix in the specific way chosen in the context of eq.~\eqref{eq:State}.

In the following, we use almost all of the same parameters as in Test Problem 1. 
Of the parameters that enter the spectral density given by eq.~(\ref{jotodomega}), we only change 
the parameter $g$ to $g=1/4$ responsible for the overall strength of the coupling, in order to be in the weak
coupling limit. 
We also set the qubit energy splitting $\varepsilon_0=1$; this parameter did not affect the evolution
in the previous example (while in the interaction picture), thus it was not set. Initial states are drawn from the physical set by sampling just as in Section~\ref{section:JC}.

Generating synthetic data is now more expensive than in the previous case. We must compute quadratures over the angular frequency $\omega$ to numerically access the correlation functions. Since these functions must be recomputed at every instance of time $t$, the full dynamics are costly to construct. To accurately generate the data within an absolute error of $10^{-6}$ at time $t=3$ using 32 grid points, we must use a frequency cutoff of $\Omega=1000$ and $2^{18}$ uniformly spaced frequency points for a numerical integration over $\omega$ using the trapezoidal rule. 

To learn each of the four correlation functions, we found it sufficient to use a $[3/3]$ Padé approximant. Empirically for this case study, we found regularization parameters of $\alpha=10^{-4}$ and $\beta=0.95$ worked sufficiently well to dampen sharp oscillations almost surely to appear without regularization. We mention in passing that we found it difficult to find a good value of $\alpha$ when $\beta=1$, and this is the main reason why we introduce the parameter $\beta$ into the regularization given by eq.~\eqref{eq:reg}.

The frame rotation \eqref{eq:corot} is part of what is learned, and the optimizer
recovers $A_{33}=i\varepsilon_0$ as
$-0.07+1.01$.

We also find that the dynamics generated by our learned correlation functions do not generalize from learning a single trajectory. Instead, we train on 30 trajectories evolved from initial conditions sampled from the space $\varrho$ of density matrices, that is, initial conditions that lead to a $\rho(0)$ with unit trace and real, positive eigenvalues.  

\begin{figure}[!htb]
\begin{centering}
\subfigure{\includegraphics[width=0.45\textwidth]{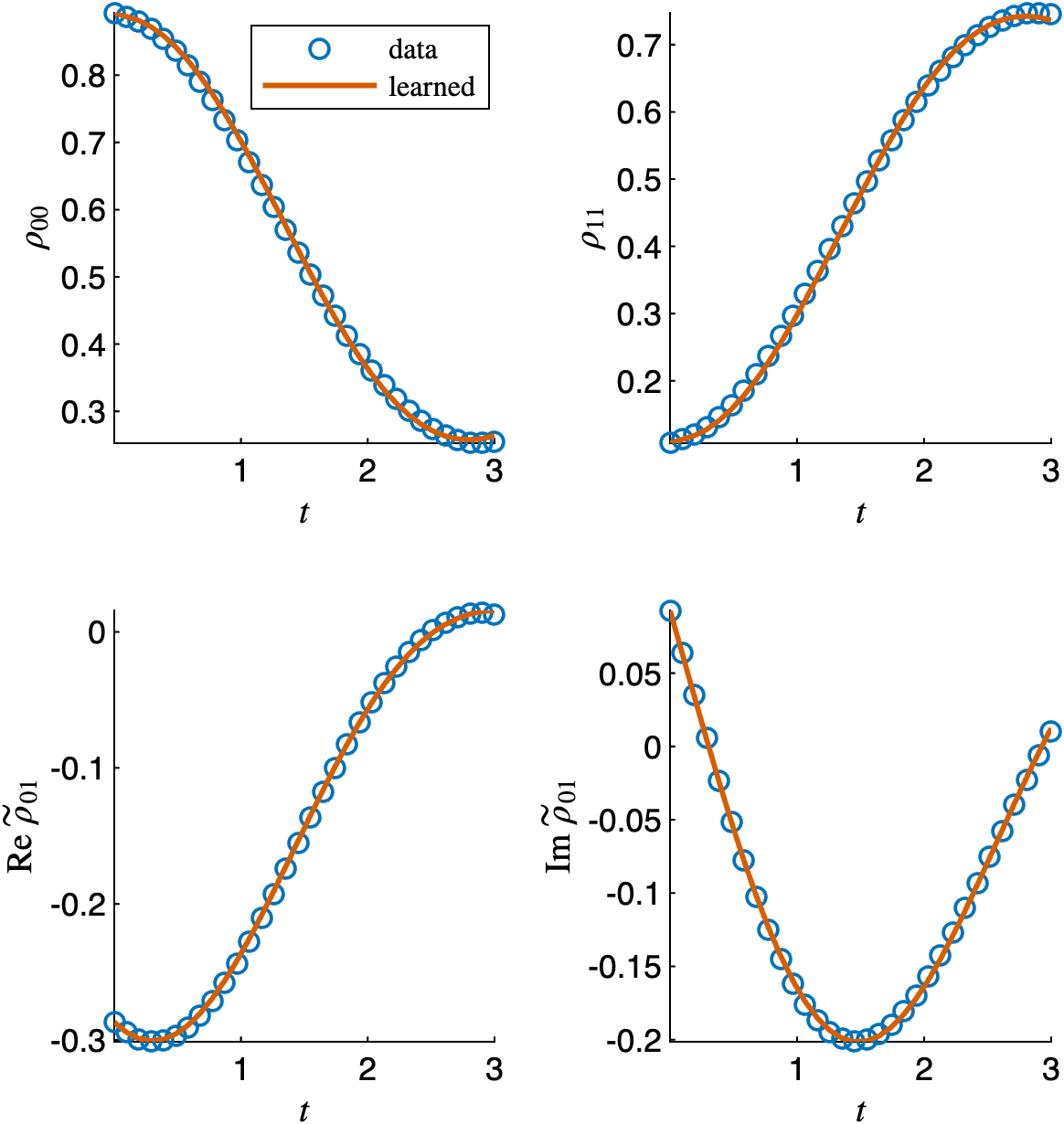}}
\subfigure{\includegraphics[width=0.45\textwidth]{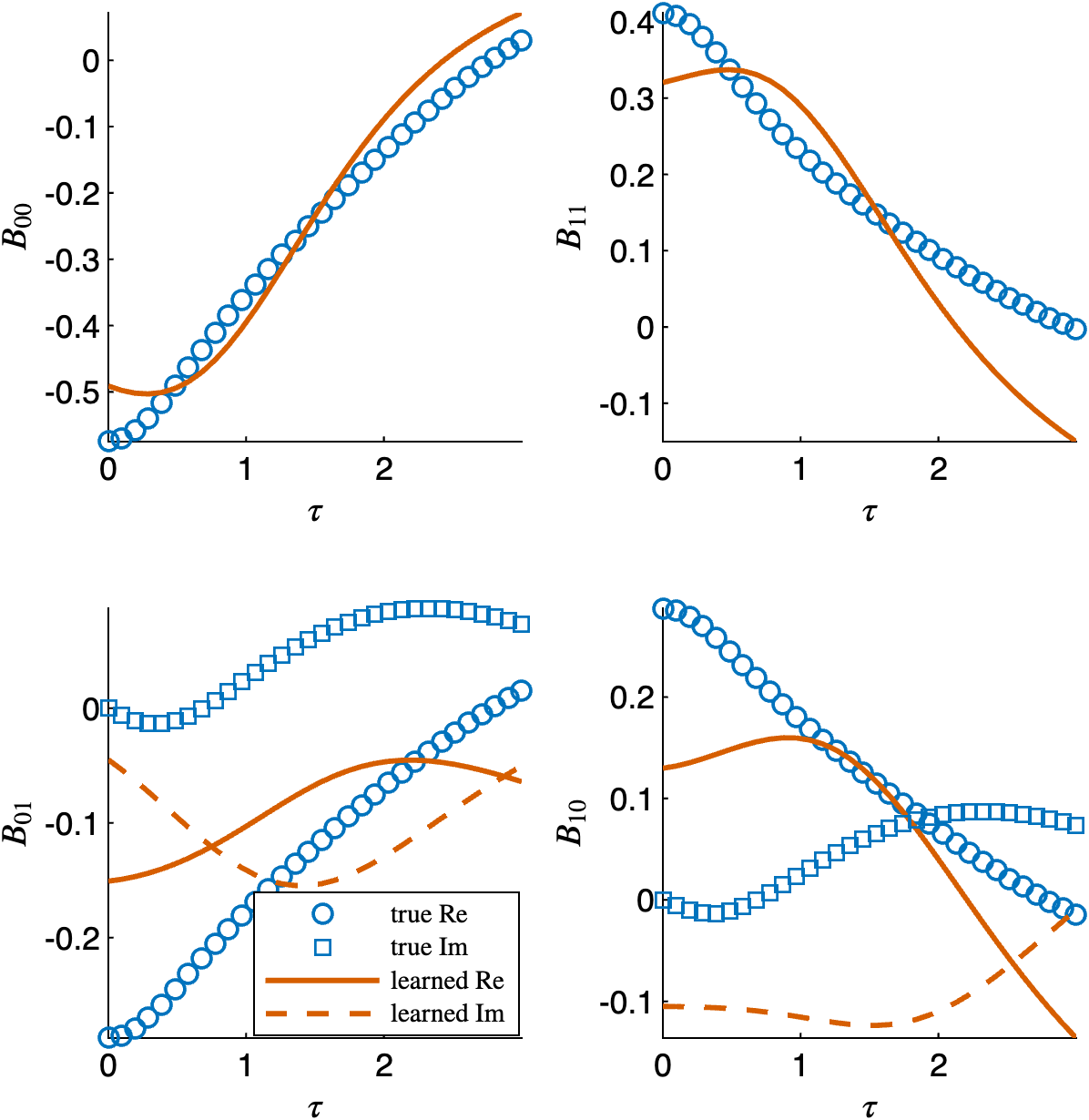}}
\end{centering}
\caption{In the top panel, we display the dynamics, governed by eqs \eqref{eq:rho-clean}, evolved from an out of training sample in the space of initial density matrices $\varrho$ and generated by the learned correlation functions in bottom panel. In the off-diagonal components, real parts correspond to open circles while imaginary parts correspond to solid circles. We observe that despite a reasonable generalization and that Tikhonov regularization ensures smoothness, the learned correlation functions are not close by any metric to the numerically evaluated correlation functions embedded in eqs~\eqref{eq:rho-clean}. 
}\label{fig:Test2}
\end{figure}

The empirical risk, given by Equation~\ref{eq:EmpRisk}, evaluated over 100 out-of-training samples is
$1.9\times10^{-4}$, the same order as the training risk $1.8\times10^{-4}$ while the minimum eigenvalue of $\rho$ over the entire ensemble remains positive. Therefore, we are left to conclude that our trained model generalizes to out of training samples and remains CP despite an inaccurate identification of the correlation function. We summarize these results visually in Figure~\ref{fig:Test2} and revisit the misidentification of the kernel in Section~\ref{sec:ident}.

\section{Test Problem 4: A qubit and a damped mode beyond the rotating-wave
approximation (quantum Rabi)}\label{section:Rabi}

The three preceding tests each supplied an analytic kernel to check against. The
same qubit-and-damped-mode apparatus of Sec.~\ref{section:JC}, with the
counter-rotating terms restored, fails to admit an exact kernel. Therefore, numerical methods for kernel identification (should one choose to model the dynamics using the Nakajima-Zwanzig route) become essential. In this section, the
ground truth is numerically accessed from an exact reduction,
helping to keep the construction of the test data provably physical.

The qubit and one bosonic mode live on $\mathcal{H}=\mathbb{C}^2\otimes\mathbb{C}^{N_{\mathrm{ph}}}$,
where the mode Fock space is truncated to $N_{\mathrm{ph}}$ levels. The Hamiltonian
restores the counter-rotating terms dropped in Sec.~\ref{section:JC},
\begin{equation}\label{eq:Rabi-ham}
H=\frac{\omega_q}{2}\,\sigma_z\otimes I
 +\omega_c\,I\otimes a^\dagger a
 +g\,\sigma_x\otimes\big(a+a^\dagger\big),
\end{equation}
and the joint state $\chi$ evolves under a Lindblad master equation with the mode
damped at rate $\kappa$,
\begin{equation}\label{eq:Rabi-lindblad}
\begin{aligned}
\dot\chi&=-i[H,\chi]+\kappa\,\mathcal{D}[a]\chi,\\
\mathcal{D}[a]\chi&=a\chi a^\dagger-\tfrac12\{a^\dagger a,\chi\}.
\end{aligned}
\end{equation}

To integrate \eqref{eq:Rabi-lindblad} we column-stack $\chi\in\mathbb{C}^{d\times d}$,
$d=2N_{\mathrm{ph}}$, into $\mathrm{vec}(\chi)\in\mathbb{C}^{d^2}$ and use the identity
$\mathrm{vec}(M\chi M')=(M'^{\top}\otimes M)\,\mathrm{vec}(\chi)$. Applied term by
term to \eqref{eq:Rabi-lindblad}, the commutator becomes
$-i(I\otimes H-H^{\top}\otimes I)$, the gain term $a\chi a^\dagger$ becomes
$\bar a\otimes a$ (since $(a^\dagger)^{\top}=\bar a$), and the two anticommutator
terms become $-\tfrac12 I\otimes a^\dagger a$ and
$-\tfrac12(a^\dagger a)^{\top}\otimes I$, so the generator is the constant matrix
\begin{equation}\label{eq:Rabi-L}
\mathcal{L}=-i\big(I\otimes H-H^{\top}\otimes I\big)
 +\kappa\Big(\bar a\otimes a-\tfrac12 I\otimes a^\dagger a
 -\tfrac12(a^\dagger a)^{\top}\otimes I\Big).
\end{equation}
Because $\mathcal{L}$ is time-independent, the exact one-step propagator is the
matrix exponential $P=e^{\mathcal{L}h}$ on the uniform grid $t_n=nh$, evaluated
once. A trajectory is the orbit $\mathrm{vec}(\chi_{n+1})=P\,\mathrm{vec}(\chi_n)$
from the factorized vacuum initial condition
$\chi_0=\rho_0\otimes|0\rangle\langle0|$, is exact to machine precision at the grid points.

The reduced qubit state is recovered at each step by the partial trace over the
mode. Writing $\chi_n$ in $2\times2$ block form with blocks of size
$N_{\mathrm{ph}}$,
\[
\chi_n=\begin{pmatrix}\chi^{(n)}_{00}&\chi^{(n)}_{01}\\[2pt]\chi^{(n)}_{10}&\chi^{(n)}_{11}\end{pmatrix},
\qquad
(\rho_n)_{ij}=\mathrm{Tr}\big(\chi^{(n)}_{ij}\big),
\]
so each reduced matrix element is the trace of one block. 
The reduced (and vectorized) state data $\{\mathbf{x}_n\}$ is then a partial trace of a completely positive
joint evolution, hence CP and trace-preserving for every coupling
strength and every truncation $N_{\mathrm{ph}}$. Initial states are drawn from the physical set, once again, as in Section~\ref{section:JC}.

We now consider the finite-dimensional operators $A$ and $B$ in this setting.
The instantaneous reduced generator is the projected Liouvillian
$A=\mathcal{P}\mathcal{L}\mathcal{P}$, where $\mathcal{P}$ is the
factorized-vacuum projector. Both the coupling and the dissipator have zero
expectation in the mode vacuum, so $A$ reduces to the bare qubit rotation,
\begin{equation}\label{eq:Rabi-A}
A=-i[H_S,\,\cdot\,]=\mathrm{diag}\big(0,\,0,\,-i\omega_q,\,+i\omega_q\big).
\end{equation}
Meanwhile, the kernel inherits sparsity from the joint parity
$\Pi=\sigma_z\otimes e^{i\pi a^\dagger a}$, which commutes with $H$ and leaves
$\mathcal{D}[a]$ invariant, and so commutes with the Nakajima--Zwanzig kernel
for any coupling strength. On the reduced qubit, $\Pi$ acts as
$\rho\mapsto\sigma_z\rho\sigma_z$, and alongside trace preservation and
Hermiticity, the admissible kernel is of the form
\begin{equation}\label{eq:Rabi-kernel}
B_{\mathrm{Rabi}}(\tau)=
\begin{pmatrix}
 b_{11} & b_{12} & 0 & 0\\
-b_{11} & -b_{12} & 0 & 0\\
 0 & 0 & b_{33} & b_{34}\\
 0 & 0 & \overline{b_{34}} & \overline{b_{33}}
\end{pmatrix}.
\end{equation}
Note that this is the same block sparsity as the transverse Born
kernel of Sec.~\ref{section:Problem2}. 

Our numerical methodology requires two further ingredients to meet the challenge of this problem. First, we observed that random multi-starting for our deployed line searches stalls. The transverse coupling lands the initial residual at $\mathcal{O}(10)$ with first-order optimality $\sim10^{4}$. To overcome this, we seed instead from transfer tensors~\cite{CerrilloCao2014} (which can also be realized through quantum process tomography in an experimental setting~\cite{ChuangNielsen1997}). The reduced dynamical map
$\mathcal{E}(t)$, defined by $\mathbf{x}(t)=\mathcal{E}(t)\mathbf{x}_0$, obeys the
same Volterra equation as the state,
\begin{equation}\label{eq:map-volterra}
\dot{\mathcal{E}}(t)=A\,\mathcal{E}(t)+\int_0^t B(t-\tau)\,\mathcal{E}(\tau)\,d\tau,
\qquad \mathcal{E}(0)=I,
\end{equation}
and we obtain $\mathcal{E}_n$ at the grid points directly, by propagating the four
basis density matrices $\{E_{11},E_{22},E_{12},E_{21}\}$ under $e^{\mathcal{L}h}$
and reading their reduced images as the columns of $\mathcal{E}_n$. Discretizing
\eqref{eq:map-volterra} with a backward difference for the derivative and a
rectangle rule for the memory integral gives a recursion that isolates the discrete
kernel at each step,
\begin{equation}\label{eq:tt-deconv}
B_n=\frac{1}{h}\left(\frac{\mathcal{E}_n-\mathcal{E}_{n-1}}{h}
   -A\,\mathcal{E}_n
   -h\sum_{m=1}^{n-1}B_{n-m}\,\mathcal{E}_m\right),
\end{equation}
where $A$ is the known local generator of Eq.~\eqref{eq:Rabi-A}.
This amounts to a causal deconvolution that returns $B_n$ from the already-known $B_1,\dots,B_{n-1}$. Then, fitting a Pad\'e entry to this discrete kernel supplies the seed. From this seed, the Levenberg--Marquardt
solver of Sec.~\ref{sec2} drops the residual
by an order of magnitude within a few iterations.

Second, we found the $H^1$ penalty could not effectively
suppress spurious behavior. We instead address this structurally by writing each
denominator as a product of decaying factors,
\begin{equation}\label{eq:factored-pole}
\mathrm{den}(\tau)=\prod_{i}\Big(1+\tau\,e^{-\xi_i}\Big),
\end{equation}
so every pole sits at $\tau=-e^{\xi_i}<0$, on the decaying side of the origin and
never in $[0,T]$, for any real $\xi_i$. The optimizer then works over the
unconstrained log-rates $\xi_i\in\mathbb{R}$, which both removes the constraint
boundary and conditions the denominator gradient, since $\partial_{\xi_i}$ acts
smoothly on $e^{-\xi_i}$.

For the numerical experiment we take $\omega_q=\omega_c=1$, $g=0.4$ (well outside
the rotating-wave regime), $\kappa=0.5$, and $N_{\mathrm{ph}}=8$, on $t\in[0,12]$
with $N_t=120$. The four independent kernel functions $b_{11},b_{12},b_{33},b_{34}$ of
Eq.~\eqref{eq:Rabi-kernel} are modeled as factored-pole $[3/3]$ Pad\'e functions
and fit over $10$ training trajectories at
$\alpha=10^{-4}$, $\beta=1$.
Direct simulation confirms the data are CP (worst training minimum eigenvalue
$5.2\times10^{-3}$).

\begin{figure}[!htb]
\begin{centering}
\subfigure{\includegraphics[width=0.45\textwidth]{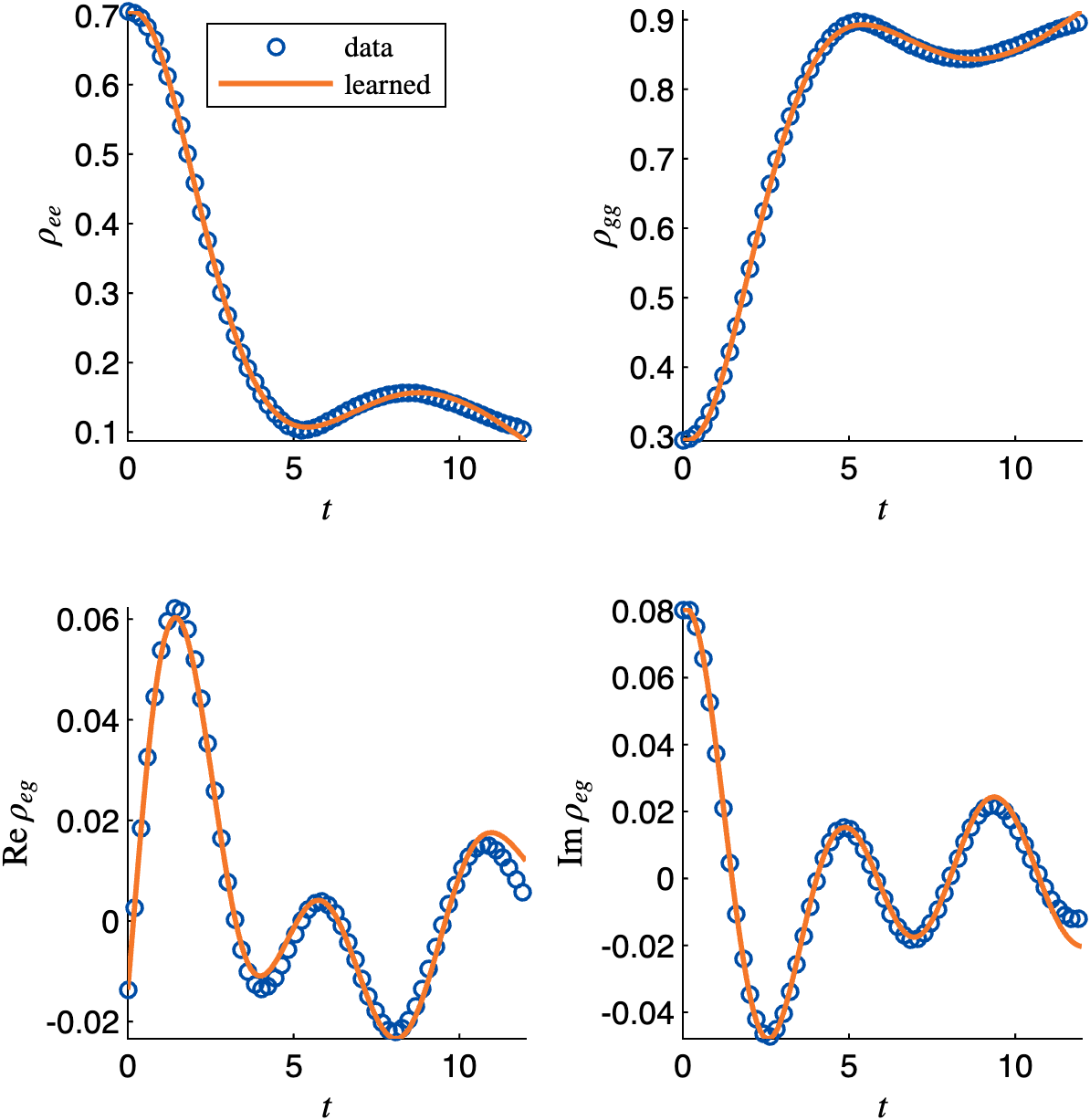}}
\subfigure{\includegraphics[width=0.45\textwidth]{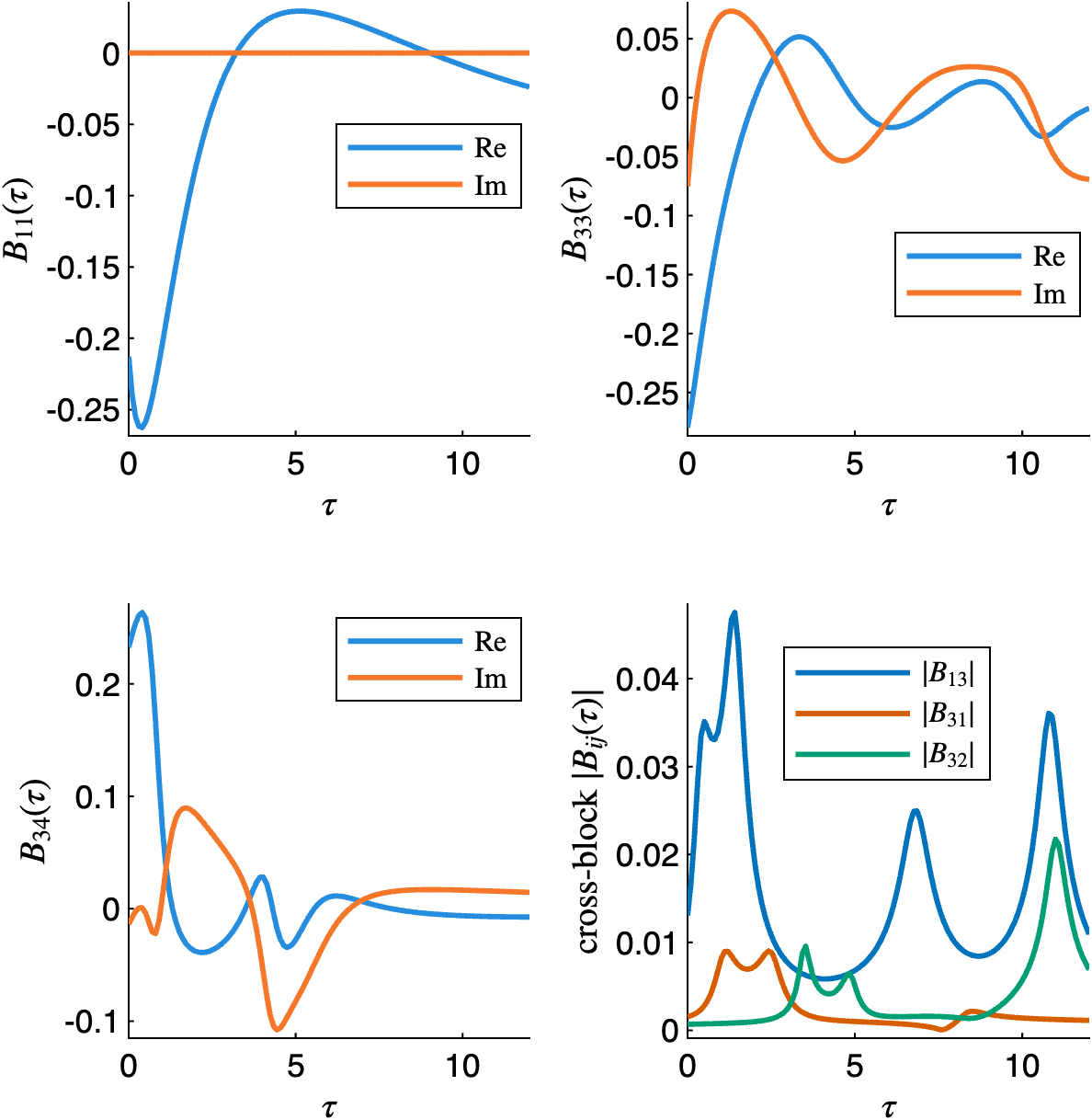}}
\end{centering}
\caption{Non-RWA Rabi test ($\omega_q=\omega_c=1$, $g=0.4$, $\kappa=0.5$,
$N_{\mathrm{ph}}=8$). Top: reduced dynamics from a held-out initial state, data
(circles) and learned model (lines). Bottom: learned kernel entries; no exact
kernel is available for this model. Training misfit $7.5\times10^{-4}$, held-out
misfit $6.8\times10^{-4}$; worst-case minimum eigenvalue
$8.2\times10^{-4}$ (learned) against $1.0\times10^{-1}$ (data).}\label{fig:Rabi}
\end{figure}

The learned model reproduces held-out trajectories to a cross-validation misfit of
$6.8\times10^{-4}$ (training $7.5\times10^{-4}$), shown with the recovered kernel in Fig.~\ref{fig:Rabi}. The learned kernel here grazes the boundary of the physical set. The worst-case minimum eigenvalue over $40$ held-out states is $8.2\times10^{-4}$, positive but
two orders of magnitude inside the data's own margin of $1.0\times10^{-1}$, whereas
the Jaynes--Cummings fit stayed level with the data. We comment that the reduced-model error is known to be mostly in the fidelity of the kernel. Against the full Fock propagation at $N_t=4096$, the solver's time-grid error at
$N_t=256$ is $7.3\times10^{-14}$ and refining $N_{\mathrm{ph}}=8\to16$ moves the
reference by $3.4\times10^{-7}$, both far below the learned relative error
$1.1\times10^{-2}$.

\section{Identifiability of the learned kernel}\label{sec:ident}

In the matrix-valued cases of Secs.~\ref{section:Problem2} and
\ref{section:Rabi}, the recovered kernels differ from the generating kernels while the state dynamics generalize. This property has also been observed in generalized Langevin and projection-operator
models~\cite{Mori1965,Zwanzig2001,Chorin2013,LiEtAl2017,JungEtAl2017}. To make quantitative statements about our numerical kernel identification here, we measure the conditioning of kernel recovery in a parametrization-invariant way, and then decompose the
realized kernel error in its sensitivity basis. This helps provide a standardized account of a kernel with an error of order one, responsible for a relative state error of order $10^{-2}$.

To begin, we write the kernel through its parameters $\theta\in\mathbb{R}^{N_\xi}$ and
stack the model output, the density-matrix trajectories over a fixed ensemble
of $K$ initial states, into one vector $\mathbf{F}(\theta)\in\mathbb{R}^{P}$,
$P=2dN_tK$. The local difficulty of
recovering $\theta$ is set by the Jacobian, computed via
\begin{equation}\label{eq:jac-def}
G_{im}=\frac{\partial F_i}{\partial \theta_m},
\qquad G\in\mathbb{R}^{P\times N_\xi},
\end{equation}
and formed by forward differences with step $10^{-6}$. To see this, use the singular value decomposition (SVD) $G=U\Sigma V^\top$, and consider 
a kernel perturbation $\delta\theta=\sum_m c_m\mathbf{v}_m$ (in the right singular basis spanned by $\{v_m\}$). The perturbation moves the
trajectories by
\begin{equation}\label{eq:fwd-pert}
\|G\,\delta\theta\|_{\ell^2(\mathbb{R}^P)}^2=\sum_m \sigma_m^2\,c_m^2.
\end{equation} 
Meanwhile, a data
error $\delta\mathbf{d}$ produces the reconstruction error via the pseudoinverse written in the SVD basis
\begin{equation}\label{eq:inv-pert}
\delta\theta=\sum_m \frac{\mathbf{u}_m^\top\delta\mathbf{d}}{\sigma_m}\,\mathbf{v}_m,
\end{equation}
We see that Eq.~\eqref{eq:fwd-pert} scales the data error by $\sigma_m$ while
Eq.~\eqref{eq:inv-pert} scales the recovered kernel error by $1/\sigma_m$ for a
data error in the same direction. Therefore, a small $\sigma_m$ means a weakly
constrained kernel and a weakly informative data direction.

However, the raw spectrum of $G$ is not intrinsic, since $\sigma_m$ measures sensitivity
per unit change of \emph{parameters}, which in principle induces an arbitrary coordinate system on kernel space. Under a smooth reparametrization $\theta=\phi(\eta)$ with coordinate
Jacobian $C=\partial\theta/\partial\eta$, both Jacobians transform as $G\mapsto
GC$ and $J\mapsto JC$, so the singular values of $G$ mix the physics of the
inverse problem with the conditioning of the Pad\'e coordinates. To remove the
coordinate dependence we measure sensitivity per unit change of the
\emph{function} $B$. 

To this end, let $\mathbf{b}(\theta)$ be the four kernel functions sampled on the
lag grid, with real and imaginary parts stacked into $8N_t$ entries. Write its
Jacobian $J_{im}=\partial b_i/\partial\theta_m$ and let $W=\mathrm{diag}(\mathbf{w})$
hold the trapezoidal quadrature weights, so that
$\mathbf{b}^\top W\,\mathbf{b}\approx\|B\|_{L^2}^2$. The parametrization-invariant sensitivities are then simply determined by the stationary values of the ratio $\|G\,\delta\theta\|_2^2/\|J\,\delta\theta\|_W^2$, interpreted as the data response per
unit $L^2$ change of the kernel, whose Euler equation is the generalized
eigenproblem~\cite{GolubVanLoan2013}
\begin{equation}\label{eq:gen-eig}
G^\top G\,\mathbf{v}_m = \lambda_m\,(J^\top W J)\,\mathbf{v}_m.
\end{equation}

The sensitivities are $\sigma_m^{\mathrm{inv}}=\sqrt{\lambda_m}$ and it's easy to see that they are the data change per unit change in $\|B\|_{L^2}$ along $\mathbf{v}_m$. Indeed, under a reparametrization $\theta=\phi(\eta)$ with $C=\partial\theta/\partial\eta$,
Eq.~\eqref{eq:gen-eig} transforms to
\begin{equation}\label{eq:gen-eig-reparam}
C^\top G^\top G\,C\,\mathbf{w}_m
   = \lambda_m\,C^\top J^\top W J\,C\,\mathbf{w}_m .
\end{equation}
The factor $C^\top(\cdot)\,C$ common to both sides cancels, so the $\lambda_m$,
and hence $\sigma_m^{\mathrm{inv}}$, are unchanged by the reparametrization.
Each $\sigma_m^{\mathrm{inv}}$ gives the change in the data per unit change in
$\|B\|_{L^2}$, as desired. See ~\cite{Rao1945} for further detail.

\begin{figure}[t]
\centering
\includegraphics[width=\columnwidth]{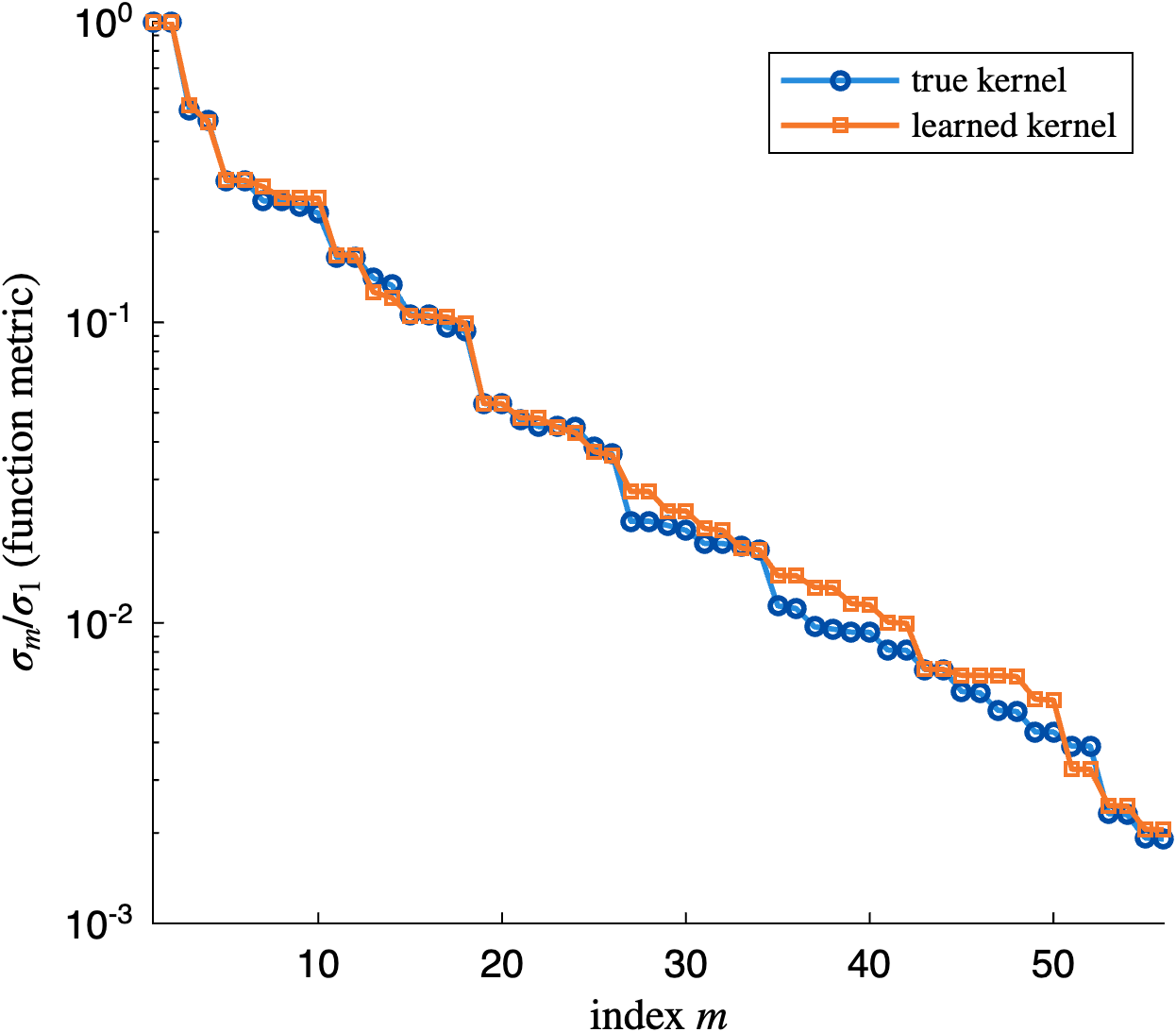}
\caption{Parametrization-invariant sensitivity spectrum
$\sigma_m^{\mathrm{inv}}/\sigma_1^{\mathrm{inv}}$ from Eq.~\eqref{eq:gen-eig},
for the transverse Born model, at the true projected kernel and at the learned
kernel in the gauge-fixed $[3/3]$ Pad\'e class. The coordinate spectrum
$\sigma_m$ (not shown) spans about twice as many orders of magnitude; the excess is Pad\'e
coordinate conditioning.}
\label{fig:ident}
\end{figure}

To evaluate spectra for the transverse Born model, we solve Eq.~\eqref{eq:gen-eig} as a dense generalized
eigenproblem. We do so on the two available kernels, the learned and the true, in the gauge-fixed $[3/3]$ Pad\'e class, with the denominator constant divided out to
remove the rational scale redundancy.  We find that the
coordinate spectrum $\sigma_m$ spans about seven orders of magnitude, while the invariant
spectra $\sigma_m^{\mathrm{inv}}$ (shown in Fig.~\ref{fig:ident}) span about three, showing the merit of using the coordinate invariant approach to isolate the native function space ill-conditioning of the inverse problem.

Let $\theta^{\mathrm{true}}$ and $\theta^{*}$ respectively denote the true and identified kernels projected onto the same
gauge-fixed $[3/3]$ class, and form
$\delta\theta = \theta^\ast - \theta^{\mathrm{true}}$. We find that $\|\delta\theta\|_2=2.24$, an order-one discrepancy consistent with the entrywise mismatch of Fig.~\ref{fig:Test2}. We build $G$ at $\theta^{\mathrm{true}}$ and take its SVD $G=U\Sigma
V^\top$ with $\sigma_1\ge\cdots\ge\sigma_{N_\xi}$, and expand
$\delta\theta=\sum_m c_m\mathbf{v}_m$ with $c_m=\mathbf{v}_m^\top\delta\theta$.
The fraction of the error captured by the first $m$ directions is
\begin{equation}\label{eq:cum-err}
E(m)=\frac{\sum_{k\le m} c_k^2}{\sum_k c_k^2},
\end{equation}
and, since direction $k$ moves the data by $\sigma_k c_k$ from
\eqref{eq:fwd-pert}, the fraction of the data response carried by the first $m$
directions is
\begin{equation}\label{eq:cum-data}
D(m)=\frac{\sum_{k\le m}\sigma_k^2 c_k^2}{\sum_k \sigma_k^2 c_k^2}.
\end{equation}

\begin{figure}[t]
\centering
\includegraphics[width=\columnwidth]{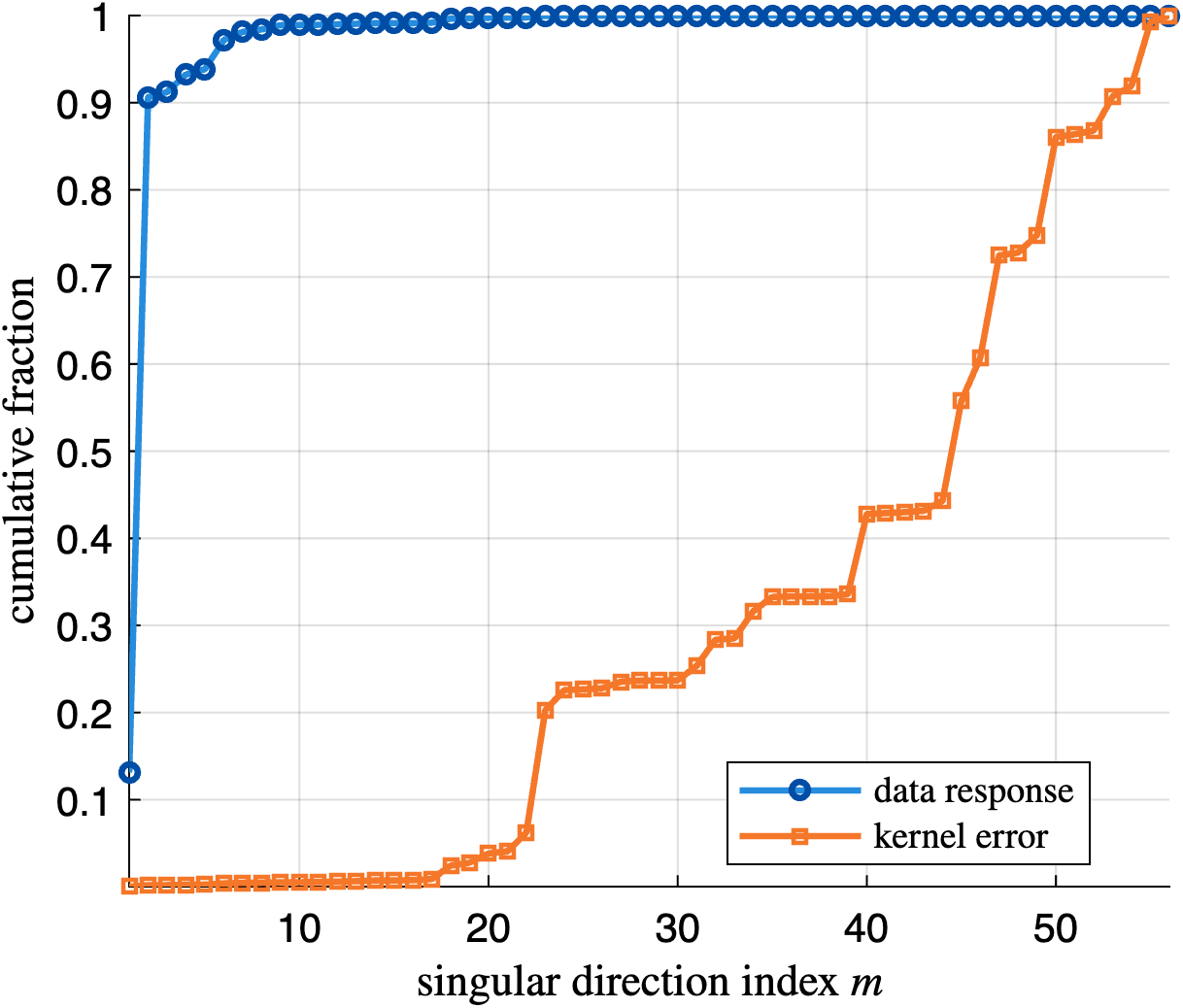}
\caption{Cumulative fraction of the data response $D(m)$ of
Eq.~\eqref{eq:cum-data} and of the kernel error $E(m)$ of
Eq.~\eqref{eq:cum-err}, for the transverse Born model, against the
singular-direction index of $G$ ordered from highest to lowest data
sensitivity. The data response saturates within the first few directions (top
two carry $90\%$), while the kernel error sits in the low-sensitivity tail (top
two carry $0.2\%$). A kernel error of order one thus produces a trajectory
error of order $10^{-2}$.}
\label{fig:ident-resolution}
\end{figure}

Ordered by decreasing data sensitivity (Fig.~\ref{fig:ident-resolution}), we find that the
top two directions carry $90\%$ of the data response and the top thirteen carry
$99\%$, while those same two hold $0.2\%$ of the kernel error and the top
thirteen hold $0.6\%$. Thresholding by conditioning instead of by count, the
directions with $\sigma_m<10^{-2}\sigma_1$ hold $96.2\%$ of the kernel error
and contribute $0.3\%$ of the data response. Thus, the error and the data response
occupy nearly disjoint subspaces.

\section{Conclusions\label{sec:conc}}
We proposed a direct, data driven route to non Markovian dynamics by learning the Volterra kernel in the Nakajima Zwanzig representation. The guiding principle was to keep the hypothesis class minimal: each scalar component of the kernel is modeled by a fixed-order Padé rational function in the lag variable, and the search is regularized by an $H^1$ penalty.  In the scalar setting the learning problem is straightforward and we find that fits are stable, trajectories are accurate, and the optimization landscape behaves predictably. Difficulties appear once we are in the matrix-valued kernel case, and require the introduction of two more ingredients, namely, transfer tensor seeding and a factored pole form of the Padé ansatz. Furthermore, non identifiability issues arise, and the behavior can no longer be attributed to the Padé model itself but rather to the structure induced by the operator-valued coupling. This point was elucidated in Section~\ref{sec:ident}.

Within this simple framework, a simple nonlocal Crank Nicolson discretization recovers accurate state trajectories across  four increasingly challenging test beds: pure dephasing, a damped Jaynes--Cummings model with an analytic coherence kernel, a transverse Born model with decoherence and population decay, and a non-rotating-wave quantum Rabi model which lacks a closed-form memory kernel.  On the analysis side, we established regularity for $B \in H^1$ and existence of minimizers for the regularized problem, which gives a basic well posedness foundation for learning operator-valued kernels.

The approach has clear limitations. First, identifiability of $(A,B)$ from finite time windows  and finite state sampling  is delicate. Distinct operator-valued kernels can generate nearly indistinguishable trajectories on $[0,T]$, and our results show that accurate state fits do not guarantee pointwise kernel recovery, a mechanism we quantify in Sec.~\ref{sec:ident}. Second, we did not enforce complete positivity. Trace preservation and Hermiticity are respected by construction, but CP is only implicit and may fail outside the training window. Third, Padé models can introduce spurious poles and local oscillations near the origin in noisy settings.  The $H^1$ penalty mitigates this behavior, and the factored-pole parametrization~\eqref{eq:factored-pole} removes it structurally by placing every denominator root off the lag window, yet, as can be seen in Figure~\ref{fig:Rabi}, sharp features in kernel recovery may still remain. Beyond larger systems, another extension is to driven qubits, where $H_S(t)$ genuinely makes the kernel two-time. The lag-only rational ansatz no longer applies, yet a separable representation in the two time arguments is the natural replacement. Finally, the nonlocal stepping with a dense history is computationally heavy, with quadratic scaling in $N_t$, which limits long horizon training and broad hyperparameter sweeps.

These observations suggest several concrete extensions. On the kernel side, barycentric rational representations~\cite{BerrutTrefethen2004,NakatsukasaSeteTrefethen2018} (or vector–fitting style constraints)  are a demonstrated improvement over Padé. The AAA fit of Table~\ref{tab:parametrization} reaches the same in-window accuracy with an order of magnitude better conditioning and avoids the spurious poles that make the Padé error non-monotone in the order.  On the structure side, one can enforce GKSL–compatible parametrizations at the memory level~\cite{Gorini1976,Lindblad1976,Vacchini2016}, e.g., by learning positive semidefinite Kossakowski tensors in a double–commutator basis. 

On the data and experimental side, longer training windows and multi–trajectory ensembles (temperatures, couplings) are suggested to help pin down kernel tails and reduce nonuniqueness~\cite{BreuerPetruccione2002,RivasHuelga2012,deVegaAlonso2017}.  We also briefly comment that because the method learns the reduced kernel from trajectories regardless of what generates them, the same pipeline applies to non-bosonic environments such as spin baths, where the bath enters only through different data.  To address this more expensive computation,  fast history evaluation via convolution–quadrature/FFT methods~\cite{Lubich1988} and sum–of–exponentials compressions reduce cost from $O(N_t^2)$ to near $O(N_t\log N_t)$; Table~\ref{tab:rabi-errors} shows the FFT route agreeing with the direct sum to machine precision while running $32\times$ to $2.2\times10^{3}$ faster across the record lengths tested, making larger grids and tighter tolerances practical. 

As a last comment, we note that yet another interpretation of non-identifiability phenomenon can be made in the frequency domain. Transforming Eq.~\eqref{eq:State}, the kernel enters through
\begin{equation}\label{eq:freq-identity}
\hat B(s)\,\hat{\mathbf{x}}(s) = (sI-A)\,\hat{\mathbf{x}}(s)-\mathbf{x}_0.
\end{equation}
It is reasonable to expect that the data constrain $\hat B$ near the transition frequencies and within the decoherence linewidth set by the relaxation rate. This is complementary to the persistency-of-excitation condition of linear system identification~\cite{Ljung1999}. Once the response has decayed, a longer time window
refines the frequencies already covered without reaching new ones. Our numerics suggest that the directions and frequencies the data leave unconstrained are those the
kernel does not act through.

The methodology of this paper returns an effective kernel, accurate where the bath shapes the qubit
dynamics and undetermined elsewhere, with the conditioning quantified by
Eq.~\eqref{eq:gen-eig} and the error localization by
Eqs.~\eqref{eq:cum-err}--\eqref{eq:cum-data}. The decomposition also gives
design guidance as follows. The high-response directions are dominated by the numerators
of the population kernels, and the unconstrained tail by the denominators of
the coherence kernels, so an efficient ansatz should spend parameters on the
former and fix or share the latter. We develop that structured parametrization (alongside the many other computational bells and whistles discussed throughout)
in follow-up work, as we look toward learning non-Markovian state dynamics in difficult scenarios that go beyond the single qubit setting.

\bigskip
\section{Acknowledgments}
J.A. acknowledges support from NSF award number 2316622.
The contribution of K.R. was supported within the QuantERA II Programme that has received funding from the EU H2020 research and innovation programme under Grant Agreement No 101017733, and with funding organisation MEYS (The Ministry of Education, Youth and Sports)
of the Czech Republic.

\appendix
\section{Numerical Method for Solving Volterra Integro-Differential Equations}\label{section:NM}

We consider the numerical solution of the Volterra-type integro–differential equation
\begin{equation}\label{eq:Volterra}
    \frac{d\mathbf{x}}{dt}
    \;=\; A\mathbf{x}(t)
    \;+\; \int_{t_0}^t B(t-\tau)\,\mathbf{x}(\tau)\,d\tau
    \;=:\;\mathbf{f}(\mathbf{x},t).
\end{equation}
This formulation permits the application of standard quadrature methods. Indeed, using the trapezoidal rule, one obtains the Crank–Nicolson update
\begin{align}
\mathbf{x}_{n+1}
  &= \mathbf{x}_n + \int_{t_n}^{t_{n+1}} \mathbf{f}(\mathbf{x},\tau)\,d\tau \notag\\
  &= \mathbf{x}_n + \tfrac{h}{2}\big(\mathbf{f}_n + \mathbf{f}_{n+1}\big) + \mathcal{O}(h^3),
\end{align}
where the time grid is uniform with step size $h:=t_{n+1}-t_n$, $\mathbf{x}(t_n)=\mathbf{x}_n$, and $\mathbf{f}(\mathbf{x}_n,t_n)=:\mathbf{f}_n$. Expanding the right-hand side gives
\begin{align}
\mathbf{f}_n + \mathbf{f}_{n+1}
   &= A\mathbf{x}_n + A\mathbf{x}_{n+1} \notag\\
   &\quad + \int_{t_0}^{t_n} B(t_n-s)\,\mathbf{x}(s)\,ds \notag\\
   &\quad + \int_{t_0}^{t_{n+1}} B(t_{n+1}-s)\,\mathbf{x}(s)\,ds .
\end{align}
Approximating these integrals once again by the trapezoidal rule and solving explicitly for $\mathbf{x}_{n+1}$ leads to the nonlocal Crank–Nicolson scheme
\begin{equation}
\mathbf{x}_{n+1}
    = \Big(I - \tfrac{h}{2}A - \tfrac{h^2}{4}B(t_0)\Big)^{-1}
      \Big(\mathbf{x}_n + \tfrac{h}{2}\,\mathbf{g}_n\Big),
\end{equation}
where the nonlocal contribution is
\begin{align}
\mathbf{g}_n
  &= A\mathbf{x}_n
     + \tfrac{h}{2}\Big(
        B(t_n)\mathbf{x}_0
        + 2\sum_{k=1}^{n-1} B(t_n-t_k)\,\mathbf{x}_k
        + B(t_0)\mathbf{x}_n \notag\\
  &\qquad\qquad
        + B(t_{n+1})\mathbf{x}_0
        + 2\sum_{k=1}^n B(t_{n+1}-t_k)\,\mathbf{x}_k
      \Big).
\end{align}
This update formula makes explicit the history dependence of the scheme: each new state $\mathbf{x}_{n+1}$ depends not only on $\mathbf{x}_n$ but on the entire trajectory $\{\mathbf{x}_k\}_{k=0}^n$ through weighted contributions of the kernel $B(\cdot)$.

\section{Computational details and validation}\label{app:details}

The forward model is the nonlocal Crank--Nicolson scheme of
Appendix~\ref{section:NM}. Integration to $N_t$ steps costs $O(d^2 N_t^2)$ per
trajectory with $d=4$, so a fit over $K$ trajectories with $N_{\mathrm{it}}$
iterations costs $O(K\,N_{\mathrm{it}}\,d^2 N_t^2)$, times $N_\xi$ for
finite-difference gradients. The $N_t^2$ dependence is removable, since the
causal history sum is a discrete convolution admitting an $O(N_t\log N_t)$ FFT
or convolution-quadrature evaluation~\cite{Lubich1988}. Scaling in dimension is set by the parameter count, that is, for a
$D$-level system,
\begin{equation}\label{eq:paramcount}
N_\xi = 2D^4(q+r+2)+2D^4 = O\!\big(D^4(q+r)\big),
\end{equation}
recovering $N_\xi=32(q+r+2)+32$ at $D=2$, so beyond one qubit the kernel must be
constrained (complete positivity, detailed balance, or a Kossakowski form), which
we note as a limitation.

The quoted losses are the unnormalized $\mathcal{J}_{\mathrm{loss}}$ of
Eq.~\eqref{eq:RegProb}. Dividing by the data energy
$\sum_j\int_0^T|x_j^{\mathrm{data}}|^2\,dt$ gives relative misfits of order
$10^{-8}$, $10^{-7}$, $10^{-5}$, $10^{-5}$ for the dephasing, Jaynes--Cummings,
transverse Born, and Rabi tests. The regularization weights come from a grid
sweep over $\alpha\in\{0,10^{-6},10^{-4},10^{-3},10^{-2}\}$,
$\beta\in\{0,0.5,0.9,0.95,1\}$, taking the plateau where the misfit is relatively insensitive
to both. We empirically found that $(10^{-6},0.5)$ for Jaynes--Cummings, $(10^{-4},0.95)$ for transverse
Born, $(10^{-4},1)$ for Rabi worked best. The scalar dephasing fit is well-conditioned and
uses $\alpha=0$.

The rational ansatz suits the Ohmic tail $\sim t^{-(p+1)}$, which a polynomial
(no decaying tail) and a finite exponential sum cannot represent compactly.
Table~\ref{tab:parametrization} fits $C(t)$ at $p=1$ by Pad\'e, Prony,
AAA~\cite{NakatsukasaSeteTrefethen2018,BerrutTrefethen2004}, and a polynomial.
We found the rational forms reach tolerance with the fewest parameters and smallest
extrapolation error, and AAA is better conditioned than Pad\'e by an order of
magnitude, marking a constrained barycentric form as the successor to interval
Pad\'e in future work.

A final comment: CP is not imposed, since it requires a positive semidefinite Choi matrix. We instead monitor $\lambda_{\min}[\rho(t)]$ for states outside the training set, equivalent for a unit-trace qubit to $\det\rho(t)=\rho_{00}\rho_{11}-|\rho_{01}|^2\ge0$. The worst-case value is
$+6.1\times10^{-5}$ for Jaynes--Cummings, $+6.6\times10^{-4}$ 
for the transverse Born model, and $+8.2\times10^{-4}$ for Rabi, all positive, so the learned dynamics stay physical without correction. Investigations into structural guarantees of CP are left to future work.

\begin{table}[t]
\centering
\setlength{\tabcolsep}{5pt}
\renewcommand{\arraystretch}{1.2}
\caption{Real parameters $N_p$ for in-window relative error $10^{-6}$, attained
in-window error, extrapolation error on $[T,2T]$, and conditioning, for four
parametrizations of the dephasing kernel $C(t)$ on $[10^{-6},3]$ at Ohmicity
$p=1$.}
\label{tab:parametrization}
\begin{tabular}{|c|c|c|c|c|}
\hline
class & $N_p$ & in-window & extrapolation & cond. number \\
\hline
Pad\'e & $13$ & $6.3\times10^{-7}$ & $4.8\times10^{-4}$ & $4.7\times10^{8}$  \\
AAA    & $13$ & $2.7\times10^{-8}$ & $8.1\times10^{-4}$ & $4.0\times10^{7}$  \\
Prony  & $28$ & $4.7\times10^{-7}$ & $5.0\times10^{-3}$ & $7.0\times10^{19}$ \\
poly.  & $19$ & $7.2\times10^{-7}$ & $3.1\times10^{8}$  & $3.0\times10^{6}$  \\
\hline
\end{tabular}
\end{table}

\begin{table}[t]
\centering
\setlength{\tabcolsep}{5pt}
\renewcommand{\arraystretch}{1.2}
\caption{Error and cost summary for the non-RWA Rabi test, with the full Fock
propagation at $N_t=4096$, $N_{\mathrm{ph}}=8$ as reference. Errors are relative
in the stacked density-matrix trajectory; the history-sum speedup is the direct
$O(N_t^2)$ wall-clock divided by the FFT $O(N_t\log N_t)$ wall-clock.}
\label{tab:rabi-errors}
\begin{tabular}{|l|c|}
\hline
\multicolumn{1}{|c|}{quantity} & value \\
\hline
full-solver grid error ($N_t=256$)            & $7.3\times10^{-14}$ \\
Fock truncation ($N_{\mathrm{ph}}=8$ vs $16$) & $3.4\times10^{-7}$  \\
learned reduced model ($N_t=256$)             & $1.1\times10^{-2}$  \\
history-sum agreement (direct vs FFT)         & $2\times10^{-12}$   \\
history-sum speedup, $N_t=256$                & $32\times$          \\
history-sum speedup, $N_t=1024$               & $550\times$         \\
history-sum speedup, $N_t=4096$               & $2200\times$        \\
\hline
\end{tabular}
\end{table}

\section{Well-posedness of the Optimization Problem}\label{section:WP}
In this section we establish a theorem that states the optimization problem used to learn the operators $A$ and $B$ is mathematically well-posed. We show that for admissible kernels $B$ the state equation admits a unique solution with sufficient regularity, and that the learning functional $\mathcal{J}$ admits at least one minimizer. The argument follows the direct method in the calculus of variations

We first establish regularity and an a priori estimate for the Volterra state equation; this will control the states associated with a minimizing sequence for the learning functional. Throughout, $\|\cdot\|_2$ denotes the Euclidean norm on $\mathbb{C}^4$, and for matrices $A\in\mathbb{C}^{4\times 4}$ we write $\|A\|$ for the spectral (matrix 2-) norm, i.e., the largest singular value of $A$.

\begin{customthm}[Lemma: Regularity and a priori estimate]\label{lem:regularity}
Let $(A,B)\in\mathcal{O}$ and consider
\begin{equation}\label{eq:State-again}
\dot{\mathbf{x}}(t)=A\,\mathbf{x}(t)+\int_{0}^{t} B(t-\tau)\,\mathbf{x}(\tau)\,d\tau,\qquad \mathbf{x}(0)=\mathbf{x}_0\in\mathbb{C}^4.
\end{equation}
If $B\in H^1\!\left(0,T;\mathbb{C}^{4\times 4}\right)$, then $\mathbf{x}\in H^1\!\left(0,T;\mathbb{C}^4\right)$ and
\begin{equation}\label{eq:apriori}
\|\mathbf{x}\|_{H^1(0,T)}\ \le\ C\!\left(\|A\|,\,\|B\|_{H^1(0,T)},\,T,\,\|\mathbf{x}_0\|_{2}\right).
\end{equation}
\end{customthm}

\begin{customproof}
First note that since $H^1(0,T)\hookrightarrow C([0,T])$, there exists a $c_T>0$ such that $\|B\|_{L^\infty(0,T)}\le c_T\|B\|_{H^1(0,T)}$. Now, for convenience, we define the Volterra operator
\begin{equation}\label{eq:KBdef} 
(\mathcal{K}_B\mathbf{y})(t):=\int_{0}^{t} B(t-\tau)\,\mathbf{y}(\tau)\,d\tau .
\end{equation}

We seek to apply Young's inequality for convolutions, which is typically stated for scalar convolutions. Since $B(\cdot)$ is matrix–valued and $x(\cdot)$ is vector–valued, we estimate pointwise using the
spectral (matrix $2$-) norm of $B$ and the Euclidean norm on $\mathbb{C}^4$:
\begin{align}\label{eq:matrix_scalar}
\|(\mathcal{K}_B x)(t)\|_{2}
&= \left\|\int_{0}^{t} B(t-\tau)\,x(\tau)\,d\tau\right\|_{2} \nonumber\\
&\le \int_{0}^{t} \|B(t-\tau)\|\,\|x(\tau)\|_{2}\,d\tau .
\end{align}
Define the scalar functions $b(s):=\|B(s)\|$ and $g(s):=\|x(s)\|_{2}$. Then
\begin{equation}\label{eq:scalar_conv}
\|(\mathcal{K}_B x)(\cdot)\|_{2} \;\le\; b * g \quad\text{on }[0,T].
\end{equation}
Thus by Young’s inequality on $[0,T]$ with $(p,q,r)=(1,2,2)$,
\begin{align}\label{eq:KB_Young} 
\|\mathcal{K}_B x\|_{L^2(0,T)}
&\le \|b*g\|_{L^2(0,T)} \nonumber\\
&\le \|b\|_{L^1(0,T)}\,\|g\|_{L^2(0,T)} \nonumber\\
&= \left(\int_0^T \|B(s)\|\,ds\right)\,\|x\|_{L^2(0,T)} \nonumber\\
&= \|B\|_{L^1(0,T)}\,\|x\|_{L^2(0,T)} .
\end{align}

Since $H^1(0,T)\hookrightarrow C([0,T])$ and from H\"older's inequality, it follows that
\begin{equation}\label{eq:Bembed}
\|B\|_{L^1(0,T)} \;\le\; \sqrt{T}\,\|B\|_{L^2(0,T)} 
\;\le\; c_T \,\|B\|_{H^1(0,T)}.
\end{equation}
Combining \eqref{eq:KB_Young} and \eqref{eq:Bembed} yields
\begin{equation}\label{eq:KBfinal}
\|\mathcal{K}_B x\|_{L^2(0,T)}
\;\le\; c_T \,\|B\|_{H^1(0,T)} \,\|x\|_{L^2(0,T)}.
\end{equation}

To estimate norms of the state vector, we use the integral formulation of the state equation \eqref{eq:State-again} which is
\begin{equation}\label{eq:integral}
\mathbf{x}(t)=\mathbf{x}_0+\int_0^t \Big(A\,\mathbf{x}(s)+(\mathcal{K}_B\mathbf{x})(s)\Big)\,ds .
\end{equation}
Taking Euclidean norms and applying \eqref{eq:KBfinal}, for $t\in[0,T]$ we obtain
\begin{align}
\|\mathbf{x}(t)\|_2
&\le \|\mathbf{x}_0\|_2
   + \int_0^t \|A\|\,\|\mathbf{x}(s)\|_2\,ds
   + \int_0^t \|(\mathcal{K}_B\mathbf{x})(s)\|_2\,ds \nonumber\\
&\le \|\mathbf{x}_0\|_2
   + \int_0^t \Big(\|A\| + c_T\|B\|_{H^1(0,T)}\Big)\,\|\mathbf{x}(s)\|_2\,ds .
\label{eq:preGronwall}
\end{align}
Finally, applying Grönwall’s inequality to \eqref{eq:preGronwall} gives the uniform bound
\begin{equation}\label{eq:Linfty}
\|\mathbf{x}\|_{L^\infty(0,T;\mathbb{C}^4)}
\le \|\mathbf{x}_0\|_2
\exp\!\Big(T\big(\|A\|+c_T\,\|B\|_{H^1(0,T)}\big)\Big).
\end{equation}

\end{customproof}

\begin{customthm}[Existence of minimizers and sequential continuity]\label{thm:wellposed}
Let $\mathcal{O}$ be as in~\eqref{eq:OperatorSet}, with $B\in H^1(0,T;\mathbb{C}^{4\times 4})$ and $A$ ranging over a nonempty, closed, and bounded subset of $\mathbb{C}^{4\times 4}$.  
Given data $\mathbf{x}^{\mathrm{data}}\in L^2(0,T;\mathbb{C}^4)$ and an initial condition $\mathbf{x}(0)=\mathbf{x}_0\in\mathbb{C}^4$, the constrained optimization problem~\eqref{eq:RegProb} admits at least one minimizer $(A_\ast,B_\ast)\in\mathcal{O}$.  

Moreover, if $(A_k,B_k)\rightharpoonup (A_\ast,B_\ast)$ in $\mathcal{O}$ with $\{(A_k,B_k)\}$ bounded, then the associated states $\mathbf{x}_k$ converge (up to subsequences) strongly in $L^2(0,T)$ to the state $\mathbf{x}_\ast$ corresponding to $(A_\ast,B_\ast)$. In particular, the loss functional $\mathcal{J}_{\mathrm{loss}}$ is sequentially continuous along bounded sequences in~$\mathcal{O}$.
\end{customthm}

\begin{customproof}Our first step is to show coercivity of the regularizer. Recall that the regularization term takes the form
\[
\mathcal{J}_{\mathrm{reg}}[B] \;=\; (1-\beta)\,\|B\|_{L^2(0,T)}^2 \;+\; \beta\,\|\dot B\|_{L^2(0,T)}^2 .
\]
Coercivity follows directly since
\begin{align}
\min\{1-\beta,\beta\}\,\|B\|_{H^1(0,T)}^2
   &\;\le\; \mathcal{J}_{\mathrm{reg}}[B] \\[0.5ex]
   &\;\le\; \max\{1-\beta,\beta\}\,\|B\|_{H^1(0,T)}^2 ,
\end{align}
implies that the sequence $\{B_k\}$ is bounded in $H^1(0,T)$, and by assumption, $\{A_k\}$ is bounded in $\mathbb{C}^{4\times 4}$.

\medskip

Now, we aim to show uniform bounds for the states and kernels. To this end, let $\mathbf{x}_k$ be the state associated with $(A_k,B_k)$.  
Lemma~\ref{lem:regularity} gives the a priori estimate
\[
\|\mathbf{x}_k\|_{H^1(0,T)} \;\le\; C\!\big(\sup_k\|A_k\|,\,\sup_k\|B_k\|_{H^1},\,T,\,\|\mathbf{x}_0\|\big).
\]
Hence $\{\mathbf{x}_k\}$ is bounded in $H^1(0,T;\mathbb{C}^4)$.  
By compactness of the embedding $H^1(0,T)\hookrightarrow L^2(0,T)$, a subsequence converges strongly:
\[
\mathbf{x}_k \;\to\; \mathbf{x}_\ast \quad \text{in } L^2(0,T).
\]
On the side of the correlation functions, since $\mathbb{C}^{4\times 4}$ is finite-dimensional, bounded sequences admit convergent subsequences: $A_k\to A_\ast$.  
For $B_k$, boundedness in $H^1(0,T)$ implies relative compactness in $C([0,T])$, so $B_k\to B_\ast$ uniformly on $[0,T]$.

\medskip

Now, we involve the dynamics. Recall that each state satisfies the integral equation
\[
\mathbf{x}_k(t) = \mathbf{x}_0 + \int_0^t \!\big(A_k \mathbf{x}_k(s) + (\Kcal_{B_k}\mathbf{x}_k)(s)\big)\,ds .
\]
For the drift term $A_k \mathbf{x}_k$, convergence follows from $A_k\to A_\ast$ and $\mathbf{x}_k\to \mathbf{x}_\ast$ in $L^2$.  
For the Volterra term, decompose
\[
\Kcal_{B_k}\mathbf{x}_k - \Kcal_{B_\ast}\mathbf{x}_\ast
= \Kcal_{B_k}(\mathbf{x}_k-\mathbf{x}_\ast) + (\Kcal_{B_k}-\Kcal_{B_\ast})\mathbf{x}_\ast .
\]
The first term vanishes by the uniform bound $\|\Kcal_{B_k}\|_{L^2\to L^2}\le c\,\|B_k\|_{H^1}$ and strong convergence of $\mathbf{x}_k$.  
The second vanishes since $B_k\to B_\ast$ uniformly, which yields $\|B_k-B_\ast\|_{L^1}\to 0$ and hence operator convergence $\Kcal_{B_k}\to \Kcal_{B_\ast}$ in $L^2\to L^2$.  
Thus the limit $\mathbf{x}_\ast$ satisfies the state equation with coefficients $(A_\ast,B_\ast)$.

\medskip

Having convergence in hand and to complete the proof, it remains to show lower semicontinuity of the functional.  But this is simple to show, by design, since the regularizer $\mathcal{J}_{\mathrm{reg}}$ is convex and weakly lower semicontinuous on $H^1(0,T)$, and 
the misfit
\[
\mathcal{J}_{\mathrm{loss}}[A,B] = \int_0^T \|\mathbf{x}^{\mathrm{data}}(t) - \mathbf{x}(t;A,B)\|^2\,dt
\]
is sequentially continuous along bounded sequences thanks to strong $L^2$ convergence of states.  Therefore,
\[
\mathcal{J}[A_\ast,B_\ast] \;\le\; \liminf_{k\to\infty} \mathcal{J}[A_k,B_k]
= \inf_{(A,B)\in\mathcal{O}} \mathcal{J}[A,B],
\]
showing $(A_\ast,B_\ast)$ is a minimizer.  
The sequential continuity claim follows from the same convergence argument.

Finally, since $\mathbf{x}\in L^\infty(0,T;\mathbb{C}^4)\subset L^2(0,T;\mathbb{C}^4)$,
the state equation \eqref{eq:State-again} gives
$\dot{\mathbf{x}}=A\mathbf{x}+\mathcal{K}_B\mathbf{x}$, so by \eqref{eq:KBfinal},
\[
\|\dot{\mathbf{x}}\|_{L^2(0,T)}
\le\big(\|A\|+c_T\|B\|_{H^1(0,T)}\big)\|\mathbf{x}\|_{L^2(0,T)}<\infty .
\]
Hence $\mathbf{x}\in H^1(0,T;\mathbb{C}^4)$ and \eqref{eq:apriori} follows.
\end{customproof}

\emph{Remark:}
The proof above relies on classical ingredients, namely, coercivity of the $H^1$–regularizer, compactness of Sobolev embeddings, Grönwall-type a priori estimates, and weak lower semicontinuity of convex functionals. These are standard tools in the analysis of Volterra equations and variational problems \cite{Brezis2010,Zeidler1990,Evans2010}. 

Nonetheless, establishing existence of minimizers and sequential continuity is essential in our setting. The learning problem is formulated over operator-valued kernels where only weak convergence of approximants is natural (e.g. when passing from Padé rational fits to their limits). Without an a priori guarantee, one could not exclude pathological minimizing sequences or loss of stability under weak convergence. Theorem~\ref{thm:wellposed} ensures that every minimizing sequence admits a subsequence converging to a genuine solution, and that the data misfit functional behaves continuously along bounded operator families. Despite its lack of mathematical novelty, this theoretical foundation justifies the numerical sections of this work and is presented here for sake of completeness.

\clearpage
\bibliography{refs}

\end{document}